\documentclass{aa} 

\usepackage[varg]{txfonts}
\usepackage{epsfig}
\usepackage{natbib}
\usepackage{amsmath}
\usepackage{amssymb}
\bibpunct{(}{)}{;}{a}{}{,} 

\usepackage{graphicx}
\usepackage[citecolor=blue,colorlinks=true,linkcolor=blue,urlcolor=blue]{hyperref}

\usepackage{siunitx}
\usepackage{bm}

\makeatletter
\renewcommand*\aa@pageof{, page \thepage{} of \pageref*{LastPage}}
\makeatother

\begin{document}

\title{Studying the multi-phase interstellar medium in the \\ Large Magellanic Cloud with \textit{SRG}/eROSITA}
\subtitle{I. Analysis of diffuse X-ray emission}

\author{Martin G.~F.~Mayer\thanks{\email{mgf.mayer@fau.de}}\inst{1} \and Manami Sasaki\inst{1} \and Frank Haberl\inst{2} \and Kisetsu Tsuge \inst{3} \and Yasuo Fukui \inst{4} \and Chandreyee Maitra\inst{2} \and Miroslav D. Filipovi\'c \inst{5} \and Zachary J.~Smeaton \inst{5} \and Lister Staveley-Smith \inst{6} \and B\"arbel Koribalski \inst{7,5} \and Sean Points \inst{8} \and Patrick Kavanagh \inst{9} }

\institute{Dr. Karl Remeis Observatory, Erlangen Centre for Astroparticle Physics, Friedrich-Alexander-Universit\"at Erlangen-N\"urnberg, Sternwartstrasse 7, 96049 Bamberg, Germany \and 
Max-Planck Institut f\"ur extraterrestrische Physik, Giessenbachstrasse, 85748 Garching, Germany \and 
Institute for Advanced Study, Gifu University, 1-1 Yanagido, Gifu, Gifu 501-1193, Japan \and 
Department of Physics, Graduate School of Science, Nagoya University, Furo-cho, Chikusa, Nagoya 464-8602, Japan \and 
Western Sydney University, Locked Bag 1797, Penrith South DC, NSW 2751, Australia \and 
International Centre for Radio Astronomy Research (ICRAR), University of Western Australia, 35 Stirling Highway, Perth, WA 6009, Australia  \and
Australia Telescope National Facility, CSIRO, Space and Astronomy, PO Box 76, Epping, NSW 1710, Australia \and
Cerro Tololo Inter-American Observatory, NOIRLab, Casilla 603, La Serena, Chile \and 
Department of Physics, Maynooth University, Maynooth, Co. Kildare, Ireland
}

\date{Received XXX /
Accepted YYY }

\abstract
{The Large Magellanic Cloud (LMC), being a nearby and actively star-forming satellite galaxy of the Milky Way, is an ideal site to observe the multi-phase interstellar medium (ISM) of a galaxy across the electromagnetic spectrum. }  
{We aim to exploit the available \textit{SRG}/eROSITA all-sky survey data to study the distribution, composition and properties of the diffuse X-ray emitting hot gas in the LMC. 
}
{We construct multi-band X-ray images of the LMC, reflecting the morphology and temperatures of the diffuse hot gas. By performing spatially resolved X-ray spectroscopy of 175 independent regions, we constrain the distribution, temperature, mass, energetics and compoistion of the hot ISM phase throughout the galaxy, while also testing for the presence of X-ray synchrotron emission.   
We combine our constraints with multiwavelength data to obtain a comprehensive view of the different ISM phases. }
{We measure a total X-ray luminosity of the hot ISM phase of $1.9\times10^{38}\,\si{erg.s^{-1}}$ ($0.2-5.0\,\si{keV}$ band), and constrain its thermal energy to around $5\times10^{54}\,\si{erg}$. The typical density and temperature of the X-ray emitting plasma are around $5\times10^{-3}\,\si{cm^{-3}}$ and $0.25\,\si{keV}$, respectively, with both exhibiting broad peaks in the southeast of the LMC.
The observed degree of X-ray absorption correlates strongly with the disitribution of foreground \ion{H}{i} gas, whereas a spatial anticorrelation between the hot and cold ISM phases is visible on sub-kpc scales within the disk. 
The abundances of light metals show a strong gradient throughout the LMC, with the north and east exhibiting a strong $\alpha$-enhancement, as expected from observed massive stellar populations there. 
In contrast, the enigmatic ``X-ray spur'' exhibits a local deficit in $\alpha$-elements, and a peak in hot-gas pressure at $P/k\sim10^5\,\si{K.cm^{-3}}$, consistent with a dominant energy input through tidally driven gas collisions.  
Finally, we tentatively identify spectroscopic signatures of nonthermal X-ray emission from the supergiant shell LMC 2, although contamination by straylight cannot be excluded. }
{}


\titlerunning{Studying the ISM in the LMC with eROSITA}
\maketitle

\section{Introduction}
The multi-phase nature of the interstellar medium (ISM) of galaxies is well established, with the first ideas of a turbulent three-phase ISM going back to \citet{Cox74} and \citet{McKee77}. In this picture, the ISM consists of a  cold ($\lesssim 10^2\,\si{K}$) and dense phase containing atomic and molecular gas and dust, in pressure equilibrium with a ``warm'' ($\sim 10^4\,\si{K}$) phase containing atomic or ionized gas, and a tenuous, hot ($\sim 10^6-10^7\,\si{K}$), and highly ionized phase, which fills a large portion of the ISM volume. While this latter phase is technically thermally unstable, cooling time-scales are typically extremely long, and energy input by supernova explosions and stellar winds sufficiently high, such that this phase does not cool appreciably within the lifetimes of massive stars \citep{Cox74}. In addition to interstellar matter, cosmic rays and magnetic fields play an important role, as their energy densities within the ISM are typically similar to the typical pressure of ordinary matter, hence impacting their galactic environment as a whole \citep[see e.g.,][]{Naab17}.  
While the cold and warm phases are optimally detected from radio (e.g., \ion{H}{I} emission at $21\,\si{cm}$) to optical (e.g., H$\alpha$ emission) wavelengths, the hot ISM phase radiates primarily in X-rays. In contrast, cosmic rays and magnetic fields in the ISM are much more challenging to observe, and can only be observed indirectly (e.g., via  diffuse nonthermal radio emission and Faraday rotation respectively).  
In any case, a multiwavelength approach is crucial for obtaining the full picture of the ISM of a galaxy, and studying the interplay of the different ISM components. 

Recent progress in the study of the cold ISM phase has been enabled, for instance, by high-resolution observations of nearby galaxies with the James Webb Space Telescope \citep{Lee23}. These have demonstrated the prevalence of voids in the cold gas, which appears to exhibit a ``swiss-cheese'' structure at low densities  due to the energy input by massive stars \citep{Sandstrom23, Barnes23}.  
However, numerous open questions persist regarding both the physics of the ISM on its own, and in the context of the  evolution of their galactic ecosystems \citep[for reviews, see][]{Ferriere01, Cox05, Naab17}. 
One example is the question of the filling factors of the different phases, in particular the hot phase, which was originally predicted to be close to volume-filling \citep{McKee77}, but may be reduced in high-density environments \citep{Girichidis16} and in the presence of magnetic fields \citep{Slavin92, Slavin93}. 
Similarly, it is unclear whether, and on what scales, the ISM is governed by pressure equilibrium between the different thermal and nonthermal phases, as global gradients in the ISM pressure of galaxies are clearly present \citep{Wolfire03}, while approximate equilibrium on small scales appears possible \citep{Korpi99}. Modern observations and simulations yield a more complex and dynamic picture than simple thermal pressure equilibrium, which involves turbulent pressure and energy exchange between different phases \citep[e.g.,][]{Koyama09, Joung09, Girichidis16,Sun20}.
Furthermore, open questions exist concerning the interplay between stellar populations and the observed ISM: regarding the hot phase, it is often suspected that a significant amount of the putative diffuse X-ray emission is in fact contributed by unresolved coronally active stars \citep[e.g.,][]{Kuntz01,Wulf19, Ponti22}. 
Similarly, while it is well-known that supernova explosions of massive stars provide significant energy input and metal enrichment to the hot ISM phase, the importance of other mechanisms which may contribute to the heating and driving of outflows in the ISM is unclear \citep[e.g.,][]{Murray11, Agertz15, Gatto17}.  
A related question is the so-called ``thermostat'' problem \citep[e.g.,][]{Cox05}, which asks how the energy injected into the ISM is mainly lost, since densities are typically too low to allow for efficient radiative cooling.
Finally, on microscopic scales, the physical conditions giving rise to diffuse X-rays in the hot ISM phase are not entirely clear. Possible scenarios include the contribution of charge-exchange emission at interfaces with colder phases \citep[e.g.,][]{Zhang22}, or the presence of non-equilibrium ionization due to recent shock-heating or rapid cooling in certain regions \citep[e.g.,][]{deAvillez12}.  

While the Milky Way, naturally, allows for studying the ISM of our own Galaxy at very high resolution, the projection effects and absorption along our sight lines complicate the study of the Galactic ISM. 
Instead, a prime target for the study of the entirety of a galaxy's ISM is the Large Magellanic Cloud (LMC). This is due to its proximity as a Milky Way satellite \citep[$d \approx 50\,\si{kpc}$,][]{distref}, relatively planar and face-on geometry \citep[$i\approx35^{\circ}$,][]{vanderMarel01}, recent star formation activity \citep[$\mathrm{SFR} \sim 0.2 \, \si{M_{\odot}.yr^{-1}}$,][]{Harris09}, and low foreground absorption from our own Galaxy. 
A large number of observing campaigns have studied the emission of the LMC across the electromagnetic spectrum. This includes surveys done in prominent optical emission lines \citep{Smith99}, near-, mid-, and far-infrared \citep{Cioni11, Meixner06, Meixner10, Meixner13}, $21\,\si{cm}$ emission from \ion{H}{I}, and in the radio continuum \citep{Tsuge24, Kim03, Pennock21}. However, in the X-ray band, the only previous imaging coverage of the entire LMC has been obtained using {\it ROSAT} \citep{Sasaki02}, while {\it XMM-Newton} has covered a significant fraction of the galaxy in mosaic pointings \citep[e.g.,][]{Maggi16, Knies21}.

Thanks to the large array of available data, the multi-phase ISM of the LMC has been studied in much detail. Observations of the cold phase via the $21\,\si{cm}$ line of \ion{H}{i} gas have shown the existence of numerous kinematically distinct components in velocity space \citep[e.g.,][]{Oh22}.  
The emission component which appears to trace the rotation of the disk (with relative velocities up to $\pm10\,\si{km.s^{-1}}$) has been labelled ``D-component''. A second component, the ``L-component'', is significantly blueshifted (by $-100$ to $-30\,\si{km.s^{-1}}$) with respect to the disk gas, and concentrated to the southeast ridge of the LMC \citep{Luks92, Fukui17, Tsuge19}.    
This is likely a manifestation of tidal interaction \citep{Fujimoto90, Oh22} between the LMC and the Small Magellanic Cloud (SMC), with the momentum exchange caused by cloud-cloud collisions \citep[see e.g.,][]{Fukui21, Tsuge19} giving rise to an intermediate \ion{H}{i} component in velocity space. Here, we define this ``I-component'' using the velocity range ($-30$ to $-10\,\si{km.s^{-1}}$) given by \citet{Tsuge19}. 
The warm ISM phase, traced for instance by optical emission lines, reveals a large number of \ion{H}{ii} regions across the LMC disk, most prominently the extremely active star-forming region 30 Doradus (30 Dor; the Tarantula Nebula) in the southeast \citep[e.g.,][]{Kennicutt88}. It appears possible that the formation of several of the observed star-forming regions in the LMC was triggered by the aforementioned collision of cold gas components \citep{Tsuge24}. 
This picture is supported also by previous X-ray observations of the southeast of the LMC. While the presence of a two-temperature plasma seems necessary to reproduce the diffuse X-ray emission across a large fraction of the LMC \citep{Sasaki21}, a larger relative amount of hot ($kT \gtrsim 0.5\,\si{keV}$) plasma appears to be present in 30 Dor and in a striking feature south of it, dubbed the ``X-ray spur'' \citep{Knies21}. Since no massive star formation appears to be ongoing in the X-ray spur, it appears possible that the dominant energy input in this region was provided by the collision of the cold gas components \citep{Knies21}. 
On larger scales, studies with the {\it ROSAT}-PSPC have demonstrated the presence of million-degree gas across the entirety of the galaxy \citep{Sasaki02}, and observations with the \textit{Einstein Observatory} have indicated an anticorrelation between hot and cold ISM phases \citep{Wang91}. However, thus far, higher spatial or spectral resolution data of the whole LMC have not been available.

\begin{figure*}[t!]
\centering
\includegraphics[width=9.15cm]{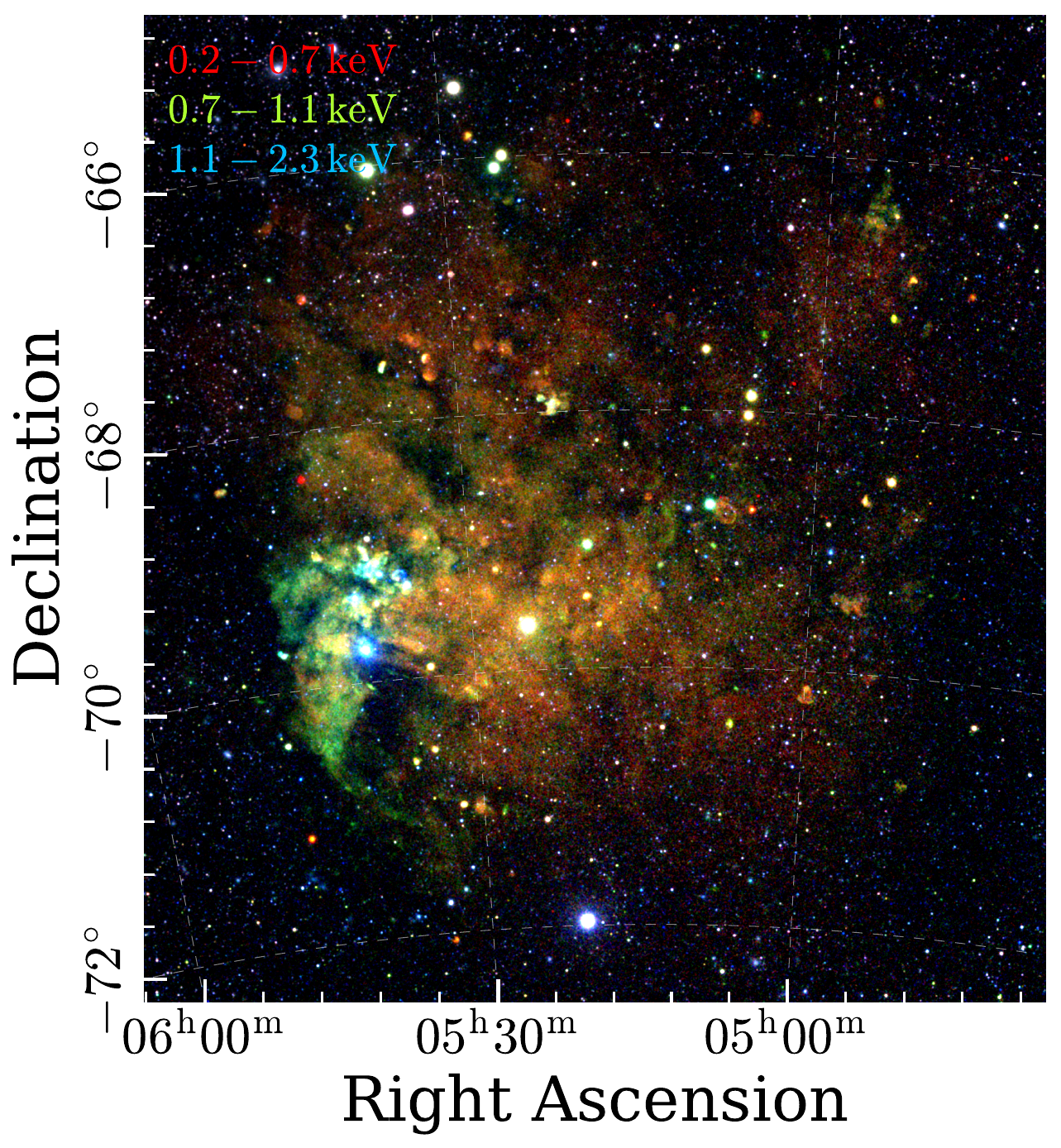} 
\includegraphics[width=9.15cm]{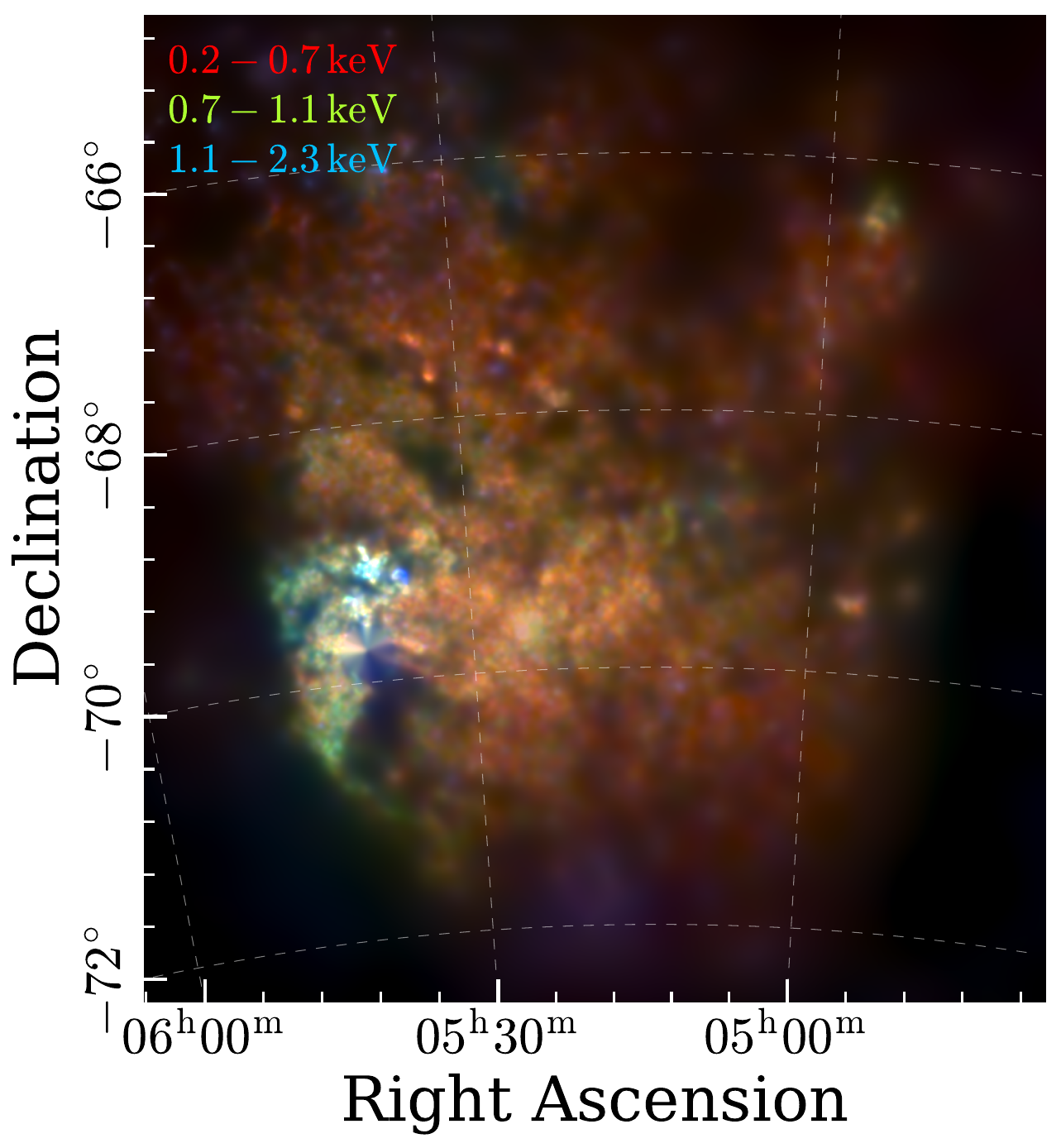} 
\caption{Exposure-corrected three-band false-color images of the LMC displayed in a square-root brightness scale. The left panel displays the exposure-corrected image without any masking of compact sources, smoothed with a Gaussian kernel of $15\arcsec$ size. 
The right panel displays the adaptively smoothed image with point-like and compact sources removed, and the resulting holes filled. } 
\label{RGBImage}
\end{figure*}

The eROSITA telescope \citep{Predehl21, Merloni12}, launched in 2019 onboard the \textit{Spectrum-Roentgen-Gamma} satellite \citep[\textit{SRG};][]{Sunyaev21}, carried out its revolutionary all-sky X-ray survey for more than two years (equivalent to around 4.5 full all-sky passages),  
its first released all-sky catalog already containing $900\,000$ X-ray sources \citep{Merloni23}. 
Owing to its privileged location close to the south ecliptic pole, serving as eROSITA survey pole, the LMC has received a comparatively large exposure in the eRASS:5 dataset.\footnote{The term ``eRASS:5'' is shorthand notation for the cumulative dataset of the (first) five eROSITA all-sky surveys, meaning the entire cumulative data set.} The total on-source time ranges between $4\,\si{ks}$ in the southwest and $30\,\si{ks}$ in the northeast of the LMC, with vignetting-corrected exposures amounting to roughly half of these respective values. 
Hence, the eRASS:5 data set provides the unprecedented opportunity to study the diffuse X-ray emitting gas across the LMC over a much larger area than available in {\it XMM-Newton} mosaics, and with a better sensitivity, spatial, and spectral resolution than achievable with {\it ROSAT}.   

This work is the first paper in a series exploiting the newly available eRASS:5 X-ray data of the LMC, in conjunction with multiwavelength information, to study the physics of the multi-phase ISM in the LMC. 
Here, we study in detail the morphological and spectroscopic properties of the X-ray emission, providing the most sensitive census of the distribution, composition, and properties of the hot X-ray emitting gas in a multiwavelength context. 
This paper is structured as follows: in Sect.~\ref{Data}, we present the data used and initial reduction steps. In Sect.~\ref{Analysis}, we describe our methods of imaging and spectroscopic analysis of X-ray and multiwavelength data, and present the resulting maps of physical quantities. In Sect.~\ref{Discussion}, we discuss the physical mechanisms giving rise to diffuse X-ray emission, study the impact of cold ISM on the distribution and emission of hot gas, consider the possibility of nonthermal X-ray emission in the southeast of the LMC, and investigate the ISM enrichment  by massive stars. Finally, we summarize our findings in Sect.~\ref{Summary}.  
Paper II (Mayer et al., in prep.) will present a more quantitative analysis of the origin of the hot ISM phase and its correlation with colder gas.

\section{Observations and data preparation\label{Data}}
\subsection{eROSITA data}
The X-ray data set we used was taken from the full eROSITA all-sky survey (eRASS:5), and was gathered between December 2019 and February 2022. We focussed on an $8^{\circ}\times8^{\circ}$ box centered on the position $(81.0^{\circ}, -68.6^{\circ})$, roughly at the center of the LMC, and compiled the necessary data set in the \texttt{c020} processing. For our entire analysis, we used the eROSITA Science Analysis Software \citep{Brunner21} in the version \texttt{eSASSusers\_211214}. 
After merging the required sky tiles \citep{Merloni23} for our region of interest using the recommended filters\footnote{\url{https://erosita.mpe.mpg.de/dr1/eSASS4DR1/eSASS4DR1_cookbook/}}, we searched and masked data affected by background flares. For this, we used the \texttt{flaregti} tool, setting a count rate threshold of $1.3\,\si{ct.s^{-1}.deg^{-2}}$ in the $4.0-8.5\,\si{keV}$ band to identify time intervals affected by flaring. 
The resulting cleaned event file contained around 22 million science-ready X-ray photons to be used as input for subsequent imaging and spectroscopic analysis.

\subsection{Multiwavelength data}
In order to obtain a complete picture of the multi-phase ISM in the LMC, we complemented our X-ray dataset with multiwavelength information, as a vast array of suitable high-quality data were readily available. 
We compiled radio continuum data from the Galactic and Extragalactic All-Sky Murchison Widefield Array Survey \citep[GLEAM,][]{Hurley17, For18} and the Australian Square Kilometer Array Pathfinder \citep[ASKAP,][]{Hotan21} early-science observations \citep{Pennock21}, to cover the distribution of cosmic rays on large and small scales, respectively. To trace the cold ISM phase, we used \ion{H}{I} $21\,\si{cm}$ observations from the Australia Telescope Compact Array (ATCA) and the 64-m Parkes telescope \citep{Kim03}, split into velocity components as in \citet{Tsuge24}. Far-, mid-, and near-infrared observations carried out within the HERITAGE and SAGE surveys \citep{Meixner10, Meixner13, Meixner06} with \textit{Herschel} and \textit{Spitzer}, respectively, trace dust of different temperatures and stellar continuum radiation. Finally, we used the images of the optical Balmer and forbidden line emission as mapped out by Magellanic Cloud Emission Line Survey \citep[MCELS;][]{Smith99}, to obtain a view of the warm phase of the ISM surrounding \ion{H}{ii} regions, planetary nebulae, and supernova remnants (SNRs).

\section{Analysis and results \label{Analysis}}
\subsection{Energy-dependent morphology \label{Imaging}}
Making use of the available sensitivity and spatial resolution of eROSITA, we created multi-band images of the X-ray emission in the LMC, to map out the distribution of the hot ISM phase. To achieve this, we used the \texttt{evtool} and \texttt{expmap} tasks \citep{Brunner21} to create count images, and vignetting-corrected exposure maps. All products were created in the three energy bands $0.2-0.7$, $0.7-1.1$, and $1.1-2.3\,\si{keV}$, as well as a total band $0.2-2.3\,\si{keV}$, and were binned to a pixel size of $15\arcsec$. 

\begin{figure}
\centering
\includegraphics[width=\linewidth]{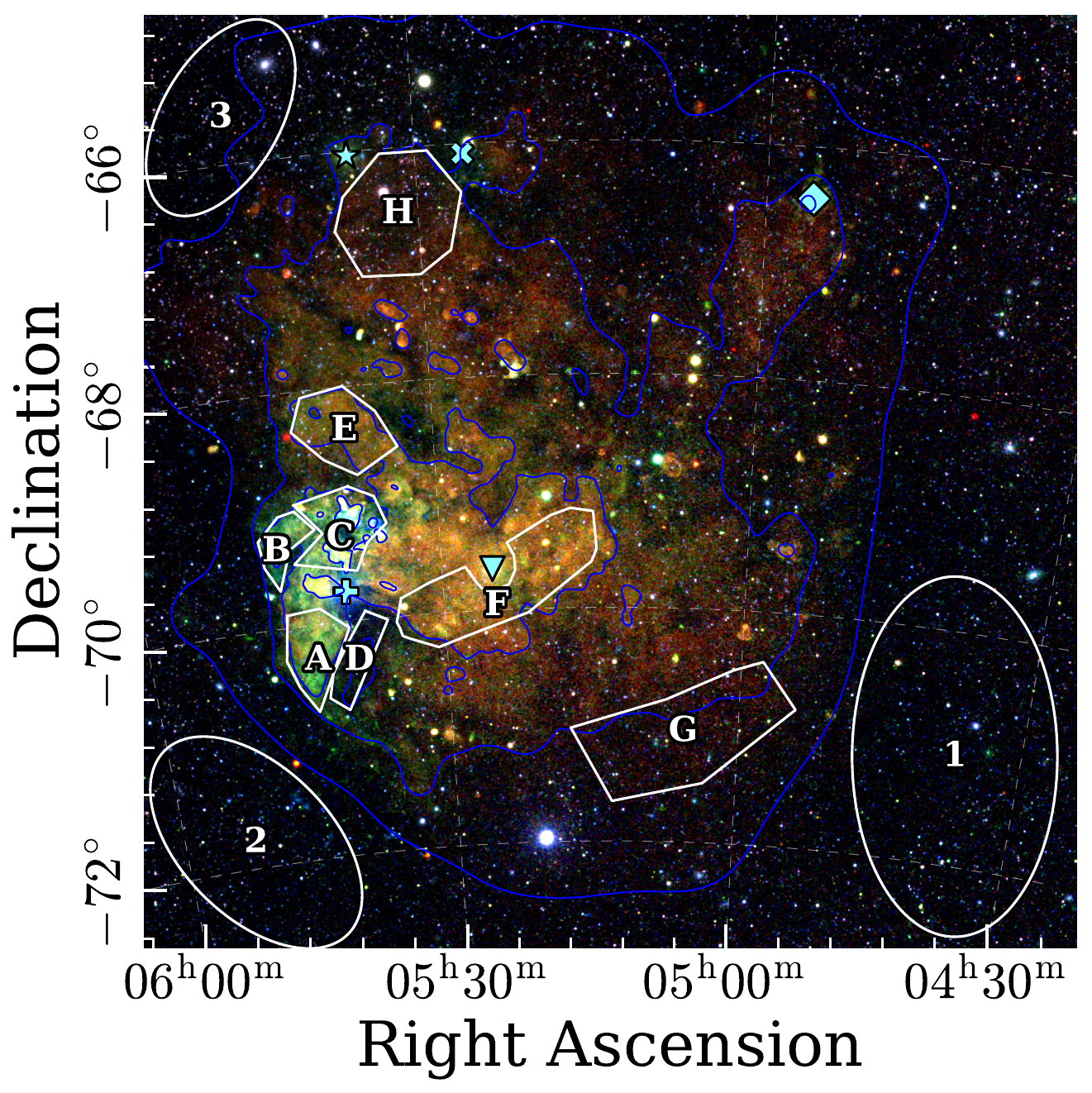} 
\caption{Map of important features and regions. We show the X-ray image of the LMC, overlaid with smoothed intensity contours (dark blue) and cyan markers indicating bright compact features: LMC X-1 (plus), SNRs N63A (star), N49 (cross), N132D (triangle), and \ion{H}{ii} region N11 (diamond). The white lines mark important extended regions, which are used for spectral extraction in Sect.~\ref{Spectroscopy}: X-ray spur (A), SGS LMC 2 (B), 30 Dor (C), X-ray dark region (D), SGS 17 (E), stellar bar (F), southern outskirts (G), and SGS LMC4 (H); The regions numbered 1 to 3 are used for background extraction. 
} 
\label{LabelImage}
\end{figure}

The resulting multi-band view of X-ray emission is displayed as a single false-color image in the left panel of Fig.~\ref{RGBImage}. A multitude of very bright point-like and compact sources dominate the image\footnote{Throughout this work, we refer to sources with a finite extent smaller than the scales relevant for diffuse emission as ``compact'', unrelated to the term  ``compact objects'' often used for black holes, neutron stars, and white dwarves.}, and for clarity we provide a labelled version of this image in Fig.~\ref{LabelImage}, which identifies the most prominent regions and features in the LMC.   
Such bright sources include high-mass X-ray binaries \citep[e.g., LMC X-1;][]{Nowak01}, and SNRs \citep[e.g., N132D;][]{Behar01} physically located inside the LMC, in addition to the population of background active galactic nuclei in the field. These respective source categories have been, and will be, treated elsewhere with eROSITA \citep[e.g.,][]{Haberl22, Maitra23, Zangrandi24}. 
In order to isolate the diffuse component of the emission from the bright sources, we proceeded in the following manner: we produced a mask based on the eRASS:4\footnote{A full-sky eRASS:5 catalog was for a long time unavailable, given the incomplete nature of the fifth eROSITA all-sky survey.} half-sky source catalog, excluding all sources with detection likelihood \citep{Brunner21} larger than 20. Around each source, we excised the radius within which its emission is expected to be above the local background level, given its count rate and the eROSITA point spread function \citep{Merloni23}. 
The resulting mask was then refined by visual inspection, excluding in particular larger regions affected by very bright and/or extended sources and known SNRs \citep[see][]{Maggi16, Zangrandi24}. The final mask was then multiplied with the respective X-ray images, and the resulting ``holes'' filled by adaptively smoothing\footnote{\url{https://xmm-tools.cosmos.esa.int/external/sas/current/doc/asmooth/index.html}} each band to $S/N=30$. In order to correct for the loss of flux in the holes, we applied the resulting smoothing template to the mask, and divided the smoothed image by the smoothed mask. 
The final smoothed and filled image, showcasing the truly diffuse emission, can be seen in the right panel of Fig.~\ref{RGBImage}. While there are no holes visible in the resulting image, unavoidable artifacts persist in the vicinity of the largest masked areas. Hence, while the presented smoothed image is highly instructive regarding the distribution of hot gas, its interpretation should be only qualitative, in particular close to the masked areas. 

Several things can be noted when inspecting the diffuse X-ray emission of the LMC: first, over wide areas, the hot ISM phase appears to be distributed rather smoothly, interrupted only by a few obvious dark areas, which may be due to either absorption or the physical absence of hot gas. 
Second, a clear gradient in the ``color'' of the emission is visible, with higher-energy emission most prominent in the south-east of the LMC, in particular in 30 Dor and the X-ray spur (labelled A and C in Fig.~\ref{LabelImage}). This may be an indication of the presence of higher-temperature gas and/or higher foreground absorption. Other regions of putatively enhanced temperatures are visible surrounding the \ion{H}{ii} region N11  (diamond marker in Fig.~\ref{LabelImage}) in the northwest \citep[see][]{Tsuge24b}, and the SNRs N63A and N49 (star, cross markers), in the north \citep{Warren03,Vancura92}. 
Finally, the X-ray emission appears to be rather sharply bounded in the south and east, whereas it seems to smoothly connect to more extended emission in the north \citep[dubbed the ``Goat Horn'',][]{Locatelli24}. At present, the nature of this extended emission complex is however unclear, as is its physical connection with the LMC. One may speculate that the sharp boundary of X-ray emission in the east could correspond to the leading edge of the galaxy moving through the Milky Way circumgalactic medium, as the LMC is moving mostly eastward with a Galactocentric velocity around $300\,\si{km\,s^{-1}}$ \citep{Gaia18c, Kallivayalil13}.  

\subsection{Spatially resolved spectroscopy \label{Spectroscopy}}
\subsubsection{Modelling approach}
The excellent spectral resolution of eROSITA \citep[$\Delta E \sim 80\,\si{eV}$ at $1.5\,\si{keV}$;][]{Predehl21} and deep exposure of the LMC region in eRASS:5 allow us to perform a systematic spectroscopic study of diffuse X-ray emission in the galaxy, to study the physical properties of the hot gas, such as temperature, density, composition, in addition to foreground absorption through intervening cold gas. 
To achieve this, we followed the established approach for spectroscopy of diffuse X-ray emission with eROSITA, previously employed in several studies of hot gas in the Galaxy and the LMC \citep{Sasaki21, Mayer22, Mayer23, Camilloni23, Yeung24}. Briefly, the region of the LMC was decomposed into regions of equal statistical weight using Voronoi tessellation, and spectra from each region were fitted using physical models of absorbed emission from hot plasma on top of physically motivated background models.

Due to the relatively weak nature of truly diffuse emission in the LMC and the significant background level, we proceeded as follows: we used the mask created in Sect.~\ref{Imaging}, to create a broad-band ($0.2-2.3\,\si{keV}$) count image, $C$, free of bright point-like and compact extended sources. This image contains a strong, but approximately uniform background contribution. To take that into account, we estimated the background level as the tenth percentile of a source-masked smoothed count-rate image of the whole field, which, when multiplied by the exposure map and source mask, yields an expectation for the background counts, $B$, in each pixel. 
Hence, we obtained an estimate of signal-to-noise per pixel of $S/N = (C-B)/\sqrt{C}$, which we fed into the adaptive Voronoi binning algorithm \citep{Vorbin, Diehl06} to create regions with an integrated $S/N = 100$. 
In total, we obtained 175 non-overlapping regions, which we used to extract spatially resolved spectra across the entire LMC. Notably, the extent of the individual regions varied more strongly with location than would be the case when not subtracting the background, due to our requirement of integrating a sufficient signal on top of the dominant background. 

We extracted spectra from each region using {\tt srctool} \citep{Brunner21}, 
limiting ourselves to data from eROSITA telescope modules 1--4 and 6, which are not affected by the optical light leak of the instrument \citep{Predehl21}. Each spectrum was then fitted in the $0.2-8.5\,\si{keV}$ range, using a composite model consisting of multi-component source emission and background templates \citep[similarly to][]{Mayer23}:
to reflect the instrumental background in our spectra, we used the well-characterized template spectrum, valid for the {\tt c020} processing, derived from filter-wheel closed data by \citet{Yeung23}.
The X-ray background model we used consisted of unabsorbed foreground components taking into account the local hot bubble and heliospheric charge exchange \citep{Yeung24}, and several components subject to Galactic foreground absorption, reflecting the Galactic halo, Galactic ``Corona'' \citep{Ponti22}, and extragalactic X-ray background \citep{DeLuca08}. In the employed spectral fitting routine {\tt Xspec} \citep{Arnaud96}, this model is expressed as {\tt acx+apec+TBabs*(apec+apec+powerlaw)}. 

Since the relative X-ray background contribution in many source regions is large, and may exhibit spatial variations across the LMC, we sampled the background spectrum in three separate large regions. These regions were spread across the outskirts of the LMC, and centered on $(\alpha, \delta) = (69.2^{\circ}, -71.1^{\circ})\,,\, (88.1^{\circ}, -71.8^{\circ})\,,\, (86.5^{\circ}, -65.6^{\circ})$, respectively (see Fig.~\ref{LabelImage}). 
The spectra, assumed to reflect the local X-ray background in each region, were fitted using our model for X-ray and instrumental background. When fitting our source spectra, all parameters but the overall background normalization were fixed, to obtain a representative template for the X-ray background in the region of interest. 
To account for potential spatial variations, the X-ray background contribution was modelled by a linear combination of the templates derived from the three regions, with the overall normalization per unit area constrained to be within a factor of two of that fitted in the background region. Thereby, the background contribution to each spectrum is more likely to be realistically represented than with a single template, and uncertainties in its shape can be accurately propagated into the uncertainties on physical parameters of the source.

As shown by previous works \citep{Knies21, Sasaki21}, the diffuse emission of the hot ISM phase in the LMC is predominantly thermal, and can be well represented by two components of optically thin plasma in collisional ionization equilibrium \citep[CIE,][]{APEC}. While typically, the LMC metallicity is considered to be around half solar, \citet{Sasaki21} showed that, close to star-forming regions, non-solar abundance ratios may be observed. Hence, we allowed varying relative abundances for typical light elements with characteristic emission lines (nitrogen, oxygen, neon, magnesium), linking their values between the two components. In order to avoid degeneracies between the overall metallicity and normalization, we kept the iron abundance fixed at $\rm Fe/H = 0.5$ \citep[relative to the ISM abundances of][]{Wilms00}. This is a sensible approximation, as interstellar iron is believed to primarily originate from type Ia supernovae \citep[e.g.,][]{Iwamoto99}, which are expected to be relatively evenly distributed throughout the LMC \citep{Maggi16}. 
In addition to the thermal components, we decided to include a power-law component in our fit, for two main reasons: first, we aimed to include the contribution to the extracted spectra of nonthermal synchrotron emission, which is expected in a few diffuse regions, most prominently 30 Dor \citep{Sasaki21}. 
Second, while the influence of point sources was mostly negated using our masking in the region definition, stray light from the extremely bright and hard source LMC X-1 \citep{Nowak01} was found to have an impact on the spectra of surrounding regions (see Sect.~\ref{NonthermalStuff}). The inclusion of a power-law component allowed us to model the contribution of this stray light to neighboring regions.

The final ingredient to our spectral model was foreground absorption by intervening cold gas, associated to both the Milky Way and the LMC. To model the Galactic component, we used the T\"ubingen-Boulder absorption model \citep{Wilms00} with the hydrogen column density given by the well-known Galactic \ion{H}{I} map presented by \citet{Dickey90}. 
The absorption intrinsic to the LMC was modelled using a separate component with half solar metallicity, and the hydrogen column was left free to vary during the fit. 
Expressed concisely, our total source model can be written as \texttt{TBabs*TBvarabs*(vapec+vapec+powerlaw)}.  

As in \citet{Mayer23}, we used the Markov Chain Monte Carlo algorithm {\tt emcee} \citep{Foreman13}, tied to the Cash statistic in {\tt Xspec}, to fit our observed spectra. We used a total of $50$ walkers, and ran the sampler for $1000$ burn-in and $2000$ sampling steps. 
For most of our quantities, we used logarithmically uniform priors. However, since we expect an absence of nonthermal emission in many fitted regions, we enforced a Gaussian prior on the power-law photon index, centered on $\Gamma=2.0$, with a width of $0.5$. This is sufficiently wide for the data to dictate the fit if a power-law-like component is present, but at the same time ensures that an upper limit can be derived for a spectrum with a realistic spectral index, in the absence of emission.      

\begin{figure*}
\centering
\includegraphics[width=18.cm]{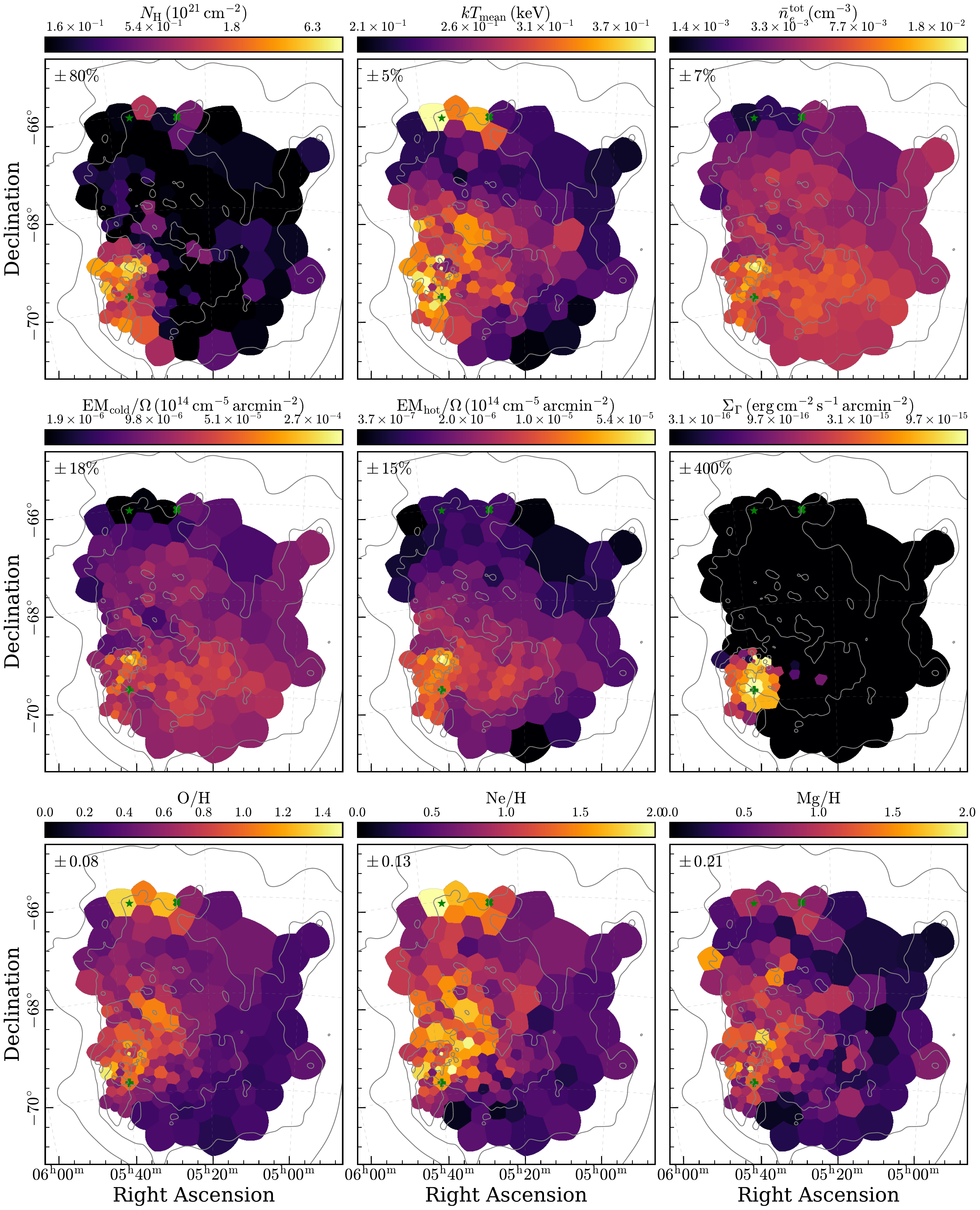} 
\caption{Maps of physical parameters derived from spectral fits to diffuse emission in the LMC. In each panel, the color map reflects the median parameter value as mapped in the respective color bar on top, while the typical (i.e., median) error is given in the upper left corner. The displayed parameters are absorption column density $N_{\rm H}$, mean temperature $kT_{\rm mean}$, electron density $\bar{n}_{e}^{\rm tot}$, emission measure per unit area $\rm EM/\Omega$, nonthermal intensity $\Sigma_{\Gamma}$, and normalized abundances of oxygen, neon, and magnesium. 
All those regions touching the edge of the image were masked in this representation, as we consider them to be background-dominated. 
The gray contours indicate the smoothed broad-band intensity of diffuse emission in the LMC, as in Fig.~\ref{LabelImage}.   
Green markers indicate bright excluded sources which may be responsible for contamination of nearby spectral extraction regions: LMC X-1 (plus), SNR N63A (star), SNR N49 (cross).} 
\label{SpecImage}
\end{figure*}

\subsubsection{Physical parameter maps}\label{ParMaps}

The results of our spectral fits are displayed in Fig.~\ref{SpecImage}, expressed in terms of physical parameters maps derived from our constraints on the model parameters. 
The top left panel displays the absorption column density $N_{\rm H}$ intrinsic to the LMC, revealing a very striking picture: While the majority of the geometric area of the LMC exhibits little to no local absorption ($N_{\rm H} \lesssim 5\times10^{20}\,\si{cm^{-2}}$), the southeast of the LMC exhibits a contiguous region of high absorption ($N_{\rm H} \sim 10^{21} - 10^{22}\,\si{cm^{-2}}$) along the rim of the galaxy. This region extends over 30 Dor, the X-ray spur, but also the X-ray dark region (region D in Fig.~\ref{LabelImage}) identified by \citet{Knies21}, partially explaining the apparent absence of X-ray emission there. 
We emphasize that the lack of significant absorption (in excess of the Galactic foreground) indicates that, on large scales and except for the southeast, the observed diffuse X-ray emission in the LMC is a very good tracer of the intrinsic distribution of the hot ISM phase.

Our fits provide a first characterization of the temperatures of the hot ISM phase across the entire LMC, which was separated into a hotter and a cooler component in our spectral model. Typically, the temperature of the cooler component was distributed quite uniformly, with the median temperature across all regions being $kT_1 = 0.21_{-0.01}^{+0.02}\,\si{keV}$ (errors reflecting the $68\%$ central interval). In contrast, the typical hotter component temperature exhibits a larger scatter, and was constrained at $kT_2 = 0.55_{-0.05}^{+0.11}\,\si{keV}$. While the cool-component temperature agrees perfectly with previous measurements, the typical temperature of the hot component is lower than in previous studies \citep{Sasaki21, Gulick21}. The reason for this is likely that the previous works were biased towards hotter gas, either because of the inclusion of unresolved thermal sources such as SNRs \citep{Gulick21}, or due to the analysis being focussed on actively star-forming regions \citep{Sasaki21}.  
In Fig.~\ref{SpecImage}, we show the distribution of the emission-weighted mean gas temperature across the LMC, which averages to $kT_{\rm mean} = 0.26_{-0.03}^{+0.06}\,\si{keV}$ over all regions. The map exhibits a clear overall gradient, with local fluctuations, showing the highest temperatures in the southeast, and typically lower temperatures in the north, west, and southwest. Interestingly, the region of enhanced temperature extends far beyond 30 Dor and the X-ray spur, the regions exhibiting the hardest X-ray emission in imaging (Fig.~\ref{RGBImage}). 
Notably, the regions surrounding the bright SNRs N63A \citep{Warren03} and N49 \citep{Vancura92} in the north appear comparatively hot. This may either be a signature of locally enhanced ISM energy input through massive star formation, or of the bright SNRs contaminating spectra extracted from their surroundings. In contrast, average temperatures fitted to the emission in 30 Dor are lower than their surroundings at $kT_{\rm mean} \approx 0.25\,\si{keV}$, presumably due to a large overdensity of cooler plasma dominating over diffuse hot gas, or due to non-equilibrium ionization mimicking a cooler component \citep{Townsley24}. 

The overall level of thermal emission is controlled by the emission measures of the two components\footnote{\url{https://heasarc.gsfc.nasa.gov/xanadu/xspec/manual/XSmodelApec.html}}, whose distribution is displayed in Fig.~\ref{SpecImage}. Both components appear to exhibit a qualitatively similar distribution, and are detected across the majority of the galaxy.
It is however noteworthy that the X-ray spur appears to exhibit a higher relative fraction of hot plasma compared to the rest of the LMC, recovering the trends observed with {\it XMM-Newton} \citep{Knies21}. 

\begin{figure*}[t!]
\centering
\resizebox{18.5cm}{!}{
\includegraphics[height=10cm]{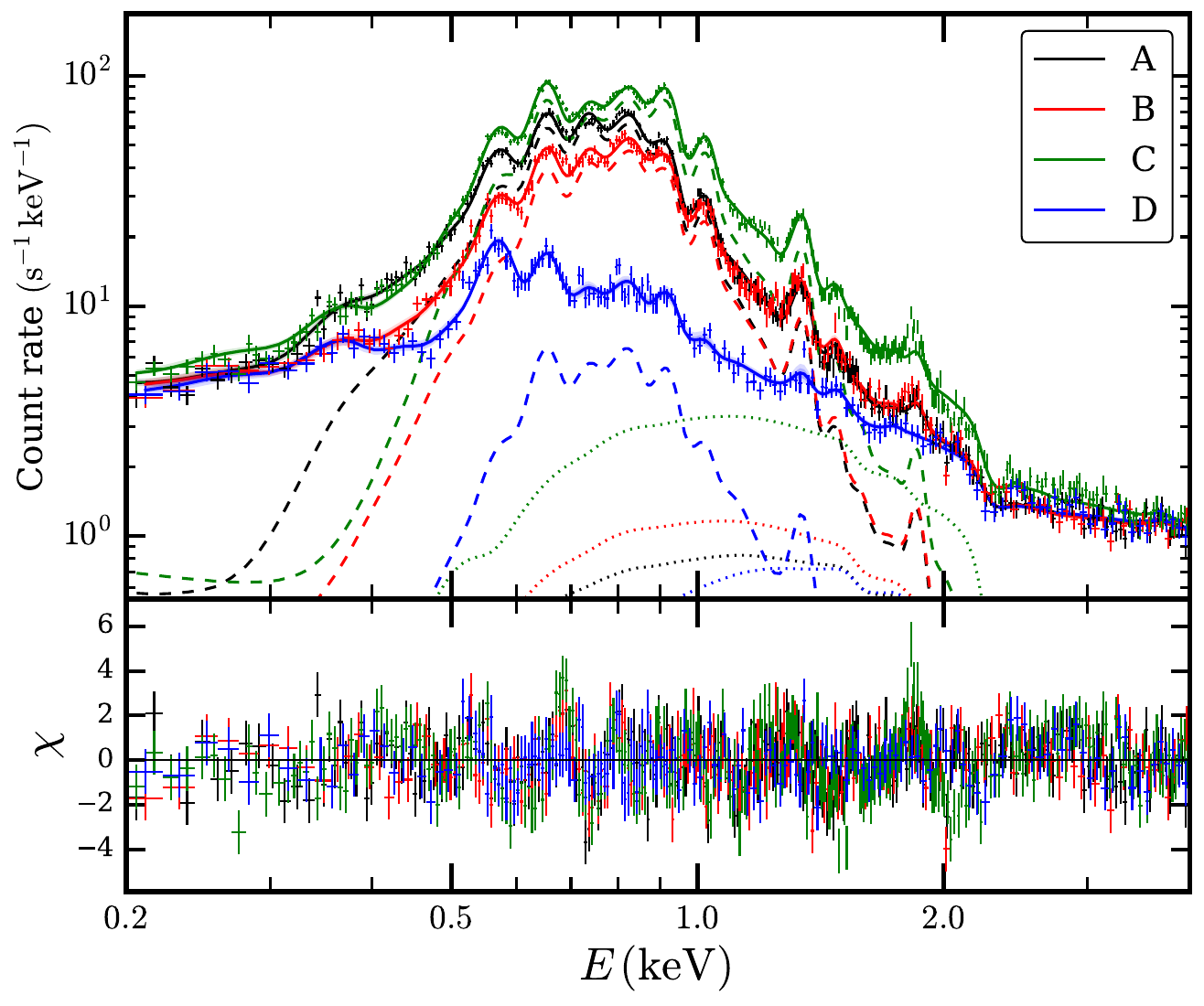} 
\includegraphics[height=10cm]{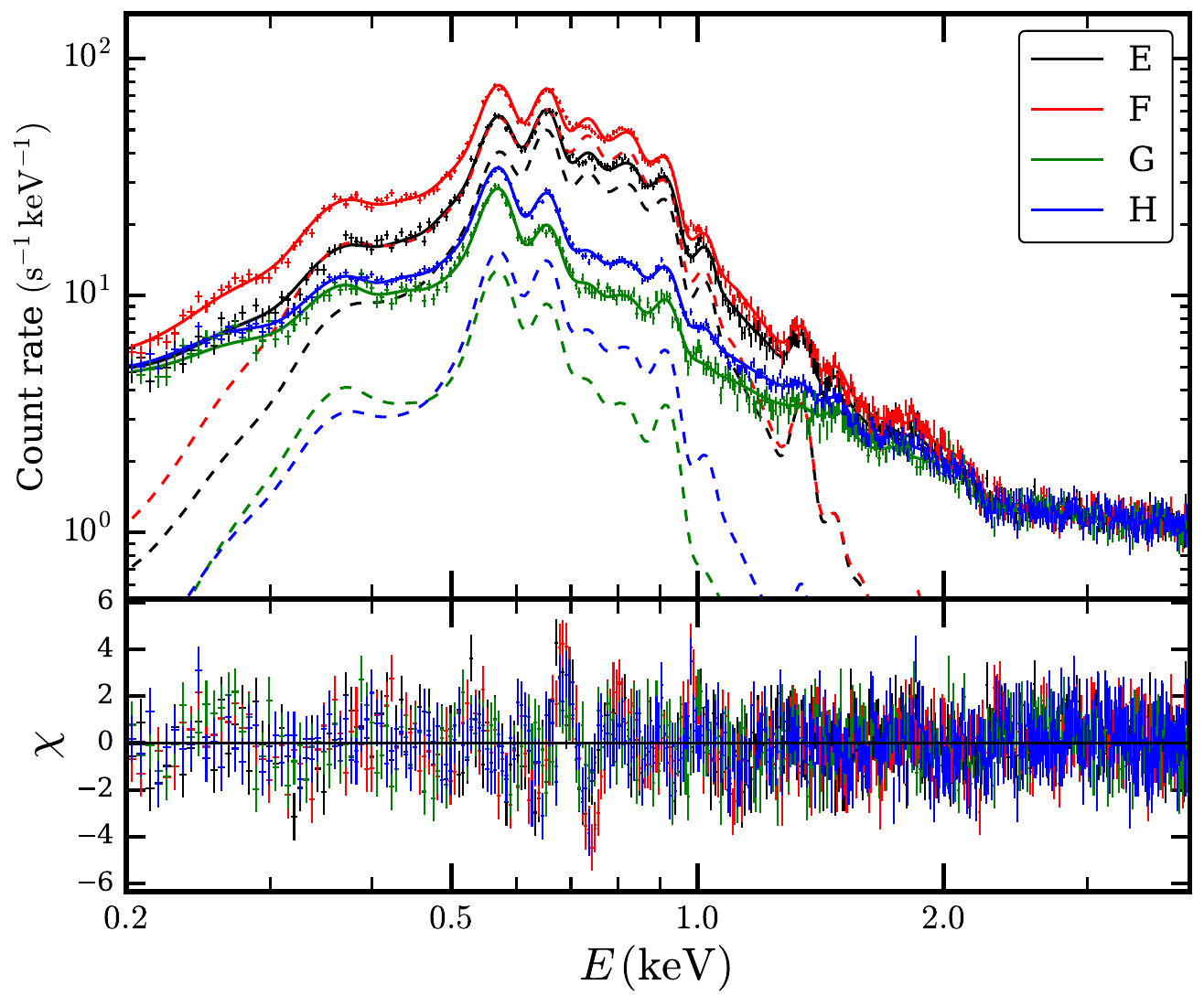} 
}
\caption{Example spectra from selected regions throughout the LMC. The two panels display the spectra of the different regions in Fig.~\ref{LabelImage} in the $0.2-4.0\,\si{keV}$ range, along with the associated error-weighted residuals in the lower sub-panels. The dashed and dotted lines correspond to the thermal and nonthermal source components, respectively. For display purposes, all spectra were normalized such that their instrumental background levels are identical.  
} 
\label{SampleSpectra}
\end{figure*}

By combining the individual components, we inferred the typical density of the hot gas as follows: we assumed that the two hot components are in approximate pressure equilibrium with each other, meaning
\begin{equation}
    n_2 = \frac{kT_1}{kT_2}n_1,
\end{equation}
where $n_{i}$ are the electron densities, and $kT_{i}$ the temperature of the fitted components, respectively. Combining this with the assumption of a constant density of each component within a region, the emission measures follow 
\begin{equation}
    \mathrm{EM}_{i} \propto n_{i}^2 \, f_{i},
\end{equation}
where $f_{i}$ is the volume filling factor of the respective component. By further defining the total filling factor of the hot phase $f = f_1+f_2$, and inserting the usual definition of the emission measure in {\tt Xspec} \citep[see also][]{Mayer22}, we obtain the following expression for the average density of the hot ISM phase:
\begin{equation}
    {n}^{\rm tot}_{e} = 10^7 \left(\frac{4\pi c_{e}}{D_{\rm LoS} f}\right)^{1/2}  \times \frac{\eta_1 kT_1+ \eta_2 kT_2}{\left(\eta_1 kT_1^2 + \eta_2 kT_2^2\right)^{1/2}},
\end{equation}
where $c_{e} \approx 1.2$ is the number of free electrons per hydrogen atom, $D_{\rm LoS}$ is the line-of-sight path length of gas through the galaxy, and we have introduced $\eta_{i} = \mathrm{EM}_{i}/\Omega$ as the normalized emission measure per unit area. 
In Fig.~\ref{SpecImage}, we show the derived average density, scaled to a unity filling factor and a line of sight of $D_{\rm LoS}=3\,\si{kpc}$, so that 
\begin{equation}
    {n}^{\rm tot}_{e} = \left ( \frac{D_{\rm LoS} f}{3\,\si{kpc}} \right )^{-1/2}\, \bar{n}^{\rm tot}_{e}.
\end{equation}
This line-of-sight depth was also used by \citet{Sasaki21}, and appears to be well-motivated given the considerable thickness of the LMC disk \citep{vdMarel02, Subramanian09}.
Considering all regions, the typical density of the hot ISM phase in the LMC is found to be $\bar{n}^{\rm tot}_{e} = 5.4^{+2.2}_{-1.7}\times10^{-3}\,\si{cm^{-3}}$.  
Assuming volume filling factors and line-of-sight depths do not change drastically, our map implies a clear density gradient, spanning about a factor of $10$ across the LMC. The highest densities of hot gas are found surrounding 30 Dor and the X-ray spur, intermediate densities coincident with the stellar bar (region F in Fig.~\ref{LabelImage}), and the lowest densities in the north of the LMC.

The bottom panels in Fig.~\ref{SpecImage} illustrate the elemental abundances of oxygen, neon, and magnesium, normalized to the solar value, across the LMC. Multiple points are noteworthy: first, the distributions of the three elements are strongly correlated to each other, arguing for a common origin of the $\alpha$-enhancement in certain regions of the ISM. The highest enrichment in $\alpha$-elements is observed in the east of the LMC, whereas the southern and western portions of the galaxy appear much poorer in those metals and may even reach subsolar $\rm \alpha/Fe$ ratios. Again, we observe an apparent enhancement surrounding the two bright SNRs in the north, which may be caused by the contamination of spectra of diffuse regions by the emission of hot metal-rich SNRs. Globally, the $\rm \alpha/Fe$ enhancement in the east of the LMC likely implies enrichment by massive stellar populations which are absent in the south and west.
The median abundances for the individual elements are given by $\rm O/H = 0.55^{+0.34}_{-0.22}$, $\rm Ne/H = 0.84^{+0.59}_{-0.36}$, and $\rm Mg/H = 0.75^{+0.37}_{-0.44}$, whereas iron was fixed at $\rm Fe/H = 0.5$. A similarly peculiar neon-to-oxygen ratio was also found in previous pointed eROSITA observations of the LMC \citep{Sasaki21}.

\begin{table*}[t!]
\renewcommand{\arraystretch}{1.5}
\caption{Physical parameters corresponding to the X-ray spectra in Fig.~\ref{SampleSpectra}. } \label{SampleSpectraPars}
\centering
\resizebox{18.4cm}{!}{
\begin{tabular}{cccccccccccc}
\hline
\hline
Feature&$N_{\rm H}$&$kT_{\rm mean}$&$\bar{n}_{e}^{\rm tot}$&$P_{\rm tot}/k$&$\epsilon$&$\log\,\tau_{\rm hot}$&$\rm O/H$&$\rm Ne/H$&$\rm Mg/H$&$\Gamma$\tablefootmark{b}&$\log\,\Sigma_{\Gamma}$\\
&$10^{21}\,\rm{cm}^{-2}$&$\rm{keV}$&$10^{-3}\,\mathrm{cm}^{-3}$&$\rm{10^3 \, K \, cm^{-3}}$&$\rm{eV \, cm^{-3}}$&&&&&$\mathrm{erg\,cm^{-2}\,s^{-1}\,arcmin^{-2}}$\\
\hline
A&$1.32_{-0.17}^{+0.18}$&$0.345_{-0.009}^{+0.008}$&$7.21_{-0.20}^{+0.26}$&$62.5_{-0.8}^{+0.9}$&$8.07_{-0.11}^{+0.12}$&&$0.60_{-0.04}^{+0.03}$&$0.97_{-0.05}^{+0.04}$&$0.84_{-0.04}^{+0.05}$&$1.0_{-0.3}^{+0.4}$&$-14.65_{-0.18}^{+0.13}$\\
B&$3.0_{-0.3}^{+0.4}$&$0.366_{-0.022}^{+0.020}$&$6.9_{-0.5}^{+0.6}$&$63.0_{-2.0}^{+2.4}$&$8.15_{-0.26}^{+0.31}$&&$0.78_{-0.08}^{+0.09}$&$1.11_{-0.09}^{+0.09}$&$0.97_{-0.08}^{+0.07}$&$1.7_{-0.4}^{+0.4}$&$-14.69_{-0.10}^{+0.08}$\\
C\tablefootmark{a}&$2.46_{-0.18}^{+0.17}$&$0.393_{-0.013}^{+0.013}$&$7.85_{-0.28}^{+0.28}$&$80.5_{-1.3}^{+1.5}$&$10.40_{-0.17}^{+0.20}$&$11.85_{-0.07}^{+0.06}$&$0.76_{-0.04}^{+0.05}$&$1.12_{-0.06}^{+0.07}$&$1.04_{-0.05}^{+0.05}$&$2.05_{-0.11}^{+0.11}$&$-14.154_{-0.030}^{+0.028}$\\
D&$5.0_{-2.1}^{+2.1}$&$0.28_{-0.06}^{+0.08}$&$4.0_{-1.4}^{+3.2}$&$28_{-5}^{+11}$&$3.6_{-0.7}^{+1.4}$&&$0.63_{-0.27}^{+0.36}$&$0.56_{-0.22}^{+0.30}$&$0.9_{-0.4}^{+0.4}$&$1.0_{-0.4}^{+0.4}$&$-14.59_{-0.10}^{+0.08}$\\ \hline
E&$0.05_{-0.03}^{+0.04}$&$0.305_{-0.006}^{+0.005}$&$4.75_{-0.08}^{+0.11}$&$37.4_{-0.6}^{+0.5}$&$4.83_{-0.08}^{+0.07}$&&$0.75_{-0.03}^{+0.03}$&$1.22_{-0.04}^{+0.05}$&$0.94_{-0.07}^{+0.07}$&$2.2_{-0.6}^{+0.6}$&$-16.20_{-0.97}^{+0.28}$\\
F&$0.022_{-0.016}^{+0.026}$&$0.2715_{-0.0030}^{+0.0022}$&$6.64_{-0.08}^{+0.12}$&$46.3_{-0.5}^{+0.5}$&$5.99_{-0.07}^{+0.07}$&&$0.450_{-0.015}^{+0.012}$&$0.787_{-0.023}^{+0.020}$&$0.60_{-0.04}^{+0.05}$&$2.1_{-0.5}^{+0.6}$&$-16.95_{-0.73}^{+0.26}$\\
G&$0.041_{-0.030}^{+0.058}$&$0.1942_{-0.0023}^{+0.0025}$&$3.99_{-0.14}^{+0.19}$&$17.1_{-0.6}^{+0.7}$&$2.21_{-0.07}^{+0.10}$&&$0.281_{-0.020}^{+0.016}$&$0.53_{-0.06}^{+0.05}$&$0.33_{-0.23}^{+0.40}$&$2.1_{-0.6}^{+0.5}$&$-17.20_{-0.61}^{+0.24}$\\
H&$0.036_{-0.026}^{+0.049}$&$0.244_{-0.005}^{+0.005}$&$2.80_{-0.10}^{+0.11}$&$17.0_{-0.5}^{+0.5}$&$2.20_{-0.06}^{+0.06}$&&$0.67_{-0.04}^{+0.04}$&$1.15_{-0.07}^{+0.06}$&$0.85_{-0.18}^{+0.16}$&$2.3_{-0.6}^{+0.7}$&$-16.66_{-0.83}^{+0.27}$\\
\hline\hline
\end{tabular}
}
\tablefoot{The fitted model and the physical parameters  correspond to those discussed in Sect.~\ref{ParMaps}. The horizontal line separates the two groups of spetra shown in separate plots in Fig.~\ref{SampleSpectra}. We show constraints on the column density $N_{\rm H}$, mean plasma temperature $kT_{\rm mean}$, electron density $\bar{n}_{e}^{\rm tot}$, energy $\epsilon$, pressure $P_{\rm tot}$, elemental abundances, as well as on nonthermal spectral index $\Gamma$ and intensity $\Sigma_{\Gamma}$. 
 \\
\tablefoottext{a}{For this region, we give the parameters, including the ionization age $\tau_{\rm hot}$, for the NEI model \texttt{vpshock} \citep{Borkowski01} in the hot component, which was strongly statistically preferred. }\tablefoottext{b}{For all regions, a Gaussian prior with mean $2.0$ and width $0.5$ was applied to $\Gamma$.}
}
\end{table*}

\begin{figure*}
\centering
\includegraphics[width=18.4cm]{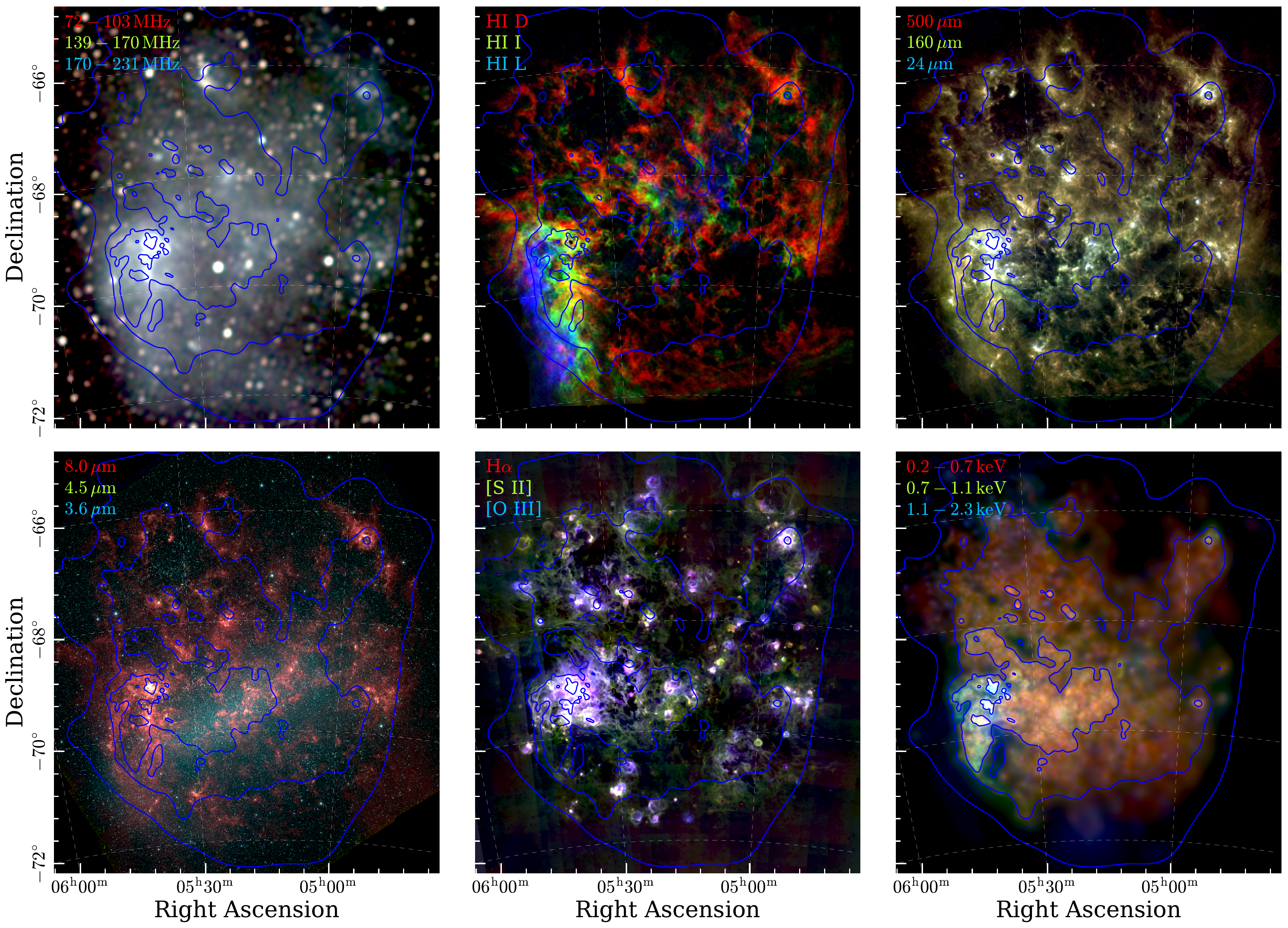} 
\caption{Multiwavelength view of the LMC. Each panel displays the emission in the region of the LMC in a different energy range. This includes the GLEAM low-frequency radio continuum (top left), ATCA and Parkes \ion{H}{i} line emission (top center), far- to mid-infrared (top right), near-infrared (bottom left), optical emission lines (bottom center), and diffuse X-rays (bottom right). 
The blue contours superimposed on each image correspond to the brightness of the diffuse X-ray emission in the $0.2-2.3\,\si{keV}$ band.
In all panels, a false-color rendering of the data using three independent energy bands is shown, with the frequency increasing from red to green to blue bands. The exception is the image of optical emission lines, where H$\alpha$, [\ion{S}{ii}], and [\ion{O}{iii}] were assigned to red, green, and blue, respectively.   
} 
\label{MultImage}
\end{figure*}

\begin{figure*}[t!]
\centering
\includegraphics[width=18.4cm]{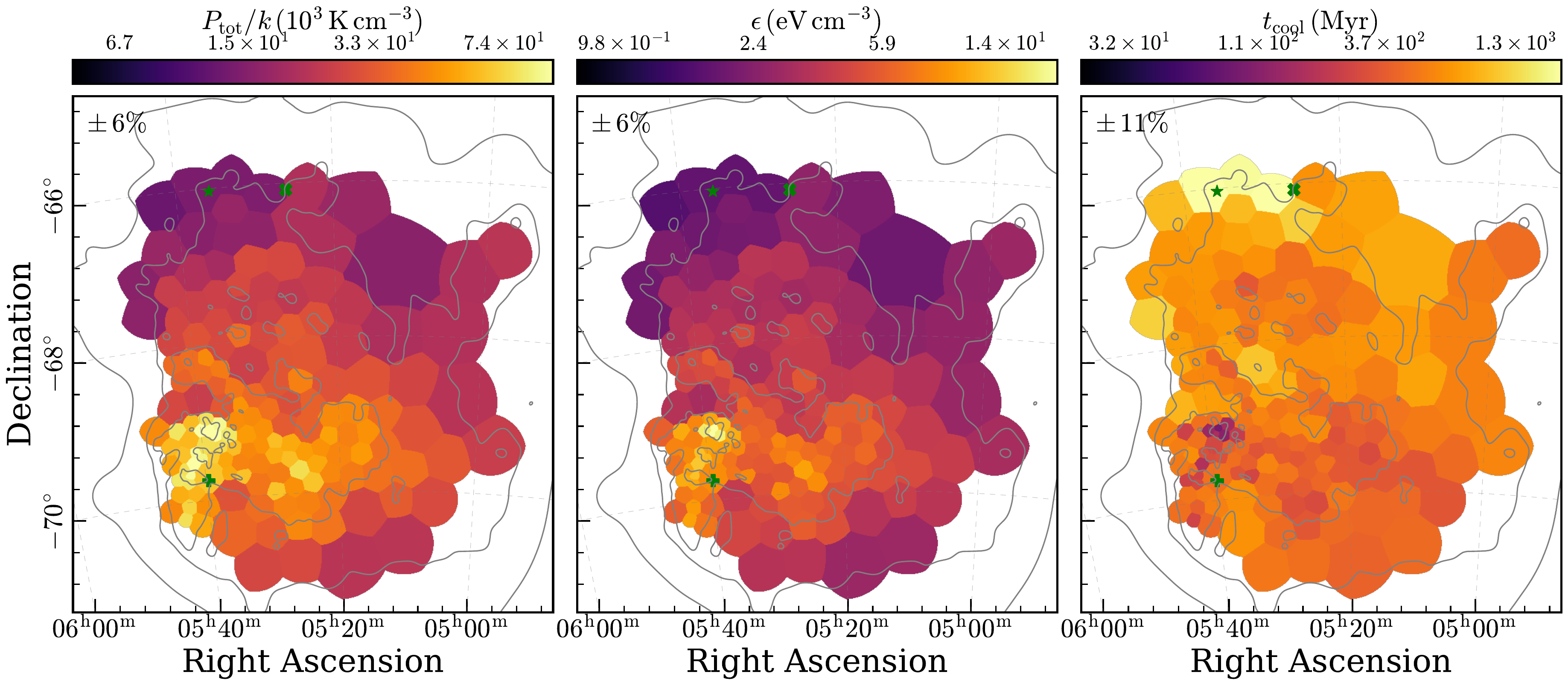} 
\caption{Distribution of thermal pressure $P/k$, energy density $\epsilon$, and cooling timescale $t_{\rm cool}$ across the LMC, derived from the physical parameters displayed in Fig.~\ref{SpecImage}.
} 
\label{EnergyPressure}
\end{figure*}

The final aspect which our fits allowed us to constrain was the level of nonthermal X-ray emission in the LMC. In Fig.~\ref{SpecImage}, we display the integrated flux of the fitted power-law component in the $1.0-5.0\,\si{keV}$ band, above the band where contamination by thermal emission is plausible. 
We do not detect nonthermal X-rays throughout the majority of the LMC, with the exception of the southeast, where surface brightnesses up to $\Sigma_{\Gamma} \sim 10^{-14}\,\si{erg.cm^{-2}.s^{-1}.arcmin^{-2}}$ are detected. While the bright synchrotron emission in the Tarantula nebula and the superbubble 30 Dor C are presumably astrophysical, the extended zone of apparent nonthermal X-ray emission surrounding LMC X-1 is likely at least partly due to contamination by bright continuum emission from the X-ray binary (see Sect.~\ref{NonthermalStuff}). 

Naturally, our analysis allows us to also constrain the properties of the LMC as an X-ray emitting galaxy as a whole. By integrating over the brightness of all source-dominated regions, we estimate the intrinsic (i.e., unabsorbed) X-ray flux of the diffuse gas in the LMC in the $0.2-5.0\,\si{keV}$ band to $F = 6.4\times 10^{-10}\,\si{erg.cm^{-2}.s^{-1}}$, corresponding to an average surface brightness of $\Sigma = 8.3\times10^{-15}\,\si{erg.cm^{-2}.s^{-1}.arcmin^{-2}}$. In combination with the well-known distance of the LMC \citep[$d\approx50\,\si{kpc}$,][]{distref}, we compute the total luminosity of the hot ISM phase to $L = 1.9\times 10^{38}\,\si{erg.s^{-1}}$. This value exceeds the luminosity of individual star-forming regions by at least an order of magnitude, including the brighest such region, 30 Dor \citep{Townsley06, Townsley24}. 
The dominant uncertainty in all of our luminosity estimates is not of statistical nature, but caused by the necessity to draw sharp boundaries around the region of interest, and by the design of our masking process, which, depending on the imposed masking level, may exclude emission from more or less bright sources. 
Our masking process is also the explanation for the large apparent disparity with the results of \citet{Gulick21}, who found an integrated flux of the LMC almost an order of magnitude higher than our value. Even though they used an almost identical spectral model to this work, their fluxes are dominated by the SNRs and point sources which we excluded, due to the non-imaging character of the \textit{HaloSat} mission. The emission attributed to thermally emitting hot gas in their work is $L = 4.9\times 10^{38}\,\si{erg.s^{-1}}$ \citep{Gulick21}, with the origin of the pronounced difference likely being the inclusion of bright SNRs. This is illustrated by the fact that the combined luminosity of some of the X-ray brightest LMC SNRs, N132D, N63A, N49, and N49B, amounts to $L = 1.8\times 10^{38}\,\si{erg.s^{-1}}$,\footnote{see \url{https://hea-www.harvard.edu/ChandraSNR/snrcat_lmc.html}} comparable to our derived value for the entire diffuse ISM. 
Similarly, the expectation for the thermal X-ray luminosity of the LMC based on its star-formation rate \citep{Harris09} and the relation of \citet{Mineo12} strongly exceeds our measurement, at $L \approx 1.5\times10^{39}\,\si{erg.s^{-1}}$. Apart from the exclusion of bright SNRs, this can additionally be explained with the systematic scatter observed around their relation, which in reality may be related to further parameters, such as metallicity or stellar mass.

Finally, we would like to discuss the adequacy of our fitted model to the observed spectra. From a statistical point of view, the goodness of fit, assessed via an estimation of the reduced $\chi^2$ statistic from rebinned spectra, is excellent, with a median value of $\chi^2_{\rm red} = 1.17_{-0.11}^{+0.17}$, and only a single region with $\chi^2_{\rm red} > 1.6$. Hence, the chosen physical model is certainly sufficiently complex to capture the behavior of the X-ray spectra of diffuse emission in the LMC.   
However, from a physical point of view, apart from its well-established nature as a model to fit diffuse emission from hot gas \citep[e.g.,][]{Knies21, Sasaki21, Gulick21, Wang91}, there are numerous alternatives to using two plasma components in CIE. The applicability of such alternative physical models is investigated quantitatively in Sect.~\ref{PhysicalNature}. 

\subsubsection{Example spectra from prominent regions} \label{ExampleSpectra}
To visualize the qualitative and quantitative variations of X-ray spectra across the LMC, in Fig.~\ref{SampleSpectra}, we display a set of eight sample spectra extracted from the regions shown in Fig.~\ref{LabelImage}, selected to reflect the range of physical properties of X-ray emitting gas across the galaxy. These include the region of the X-ray spur (labelled A), supergiant shell (SGS) LMC 2 (B), 30 Dor (C), and strongly absorbed ``hole'' in X-ray emission (D), as well as a peak coincident with SGS 17 \citep[E;][]{Kim99}, and regions aimed at the highest concentration of stars in the stellar bar (F), the southern ``outskirts'' (G) and SGS LMC 4 in the north \citep[H;][]{Meaburn80}.  
All spectra were fitted with models equivalent to the ones used to construct our parameter maps in Sect.~\ref{ParMaps}, and the best fits and corresponding parameters are shown in Fig.~\ref{SampleSpectra} and Table \ref{SampleSpectraPars}.  

Our fits clearly recover the strong foreground absorption in the southeast, with the remaining regions appearing essentially unabsorbed. Interestingly, while the X-ray hole (region D) does exhibit the strongest absorption ($N_{\rm H} \sim 5\times10^{21}\,\si{cm^{-2}}$), it also exhibits a much lower (energy) density than the neighboring X-ray spur, showcasing that the peculiar shape of the spur is not an artifact of absorption, but in fact traces the distribution of X-ray emitting gas in the region.
The highest densities, temperatures, and pressures in the hot ISM phase are clearly present in regions A, B, C, with the important distinction that region C (30 Dor) strongly prefers a model with non-equilibrium ionization (NEI) in the hot component, while the other spectra are fitted adequately with two CIE plasmas. Correspondingly, the regions furthest away from 30 Dor (G and H) exhibit the coolest, least dense, and lowest-pressure plasma.
The largest elemental abundances appear in regions B, C, E, H, which intriguingly either contain massive star-forming regions (30 Dor), or coincide with supergiant shells, arguing in favor of enrichment by massive stars. While marginal, the X-ray spur (region A) exhibits somewhat lower metal abundances, and regions F (the stellar bar) and G appear the least enriched. 
Furthermore, it is noteworthy that, at energies $\lesssim0.5\,\si{keV}$, the X-ray emission from the bar is by far the brightest of all regions, which one might ascribe to the contribution of a large amount of unresolved low-mass stars \citep[similarly to][]{Wulf19}.
Finally, ``nonthermal'' emission is clearly detected in all regions close to LMC X-1 (A, B, C, D), with flat spectral indices arguing in favor of contamination by the X-ray binary in regions A, D, and possibly B (see Sect.~\ref{NonthermalStuff}).

\subsection{A multiwavelength view of ISM in the LMC}
In this section, we compile a comprehensive multiwavelength view of the LMC from radio to X-rays, in order to reveal cosmic rays, dust, and gas of all temperatures. Apart from our smoothed X-ray images, we used archival images covering the radio continuum emission in the ranges $72-103$, $139-170$, and $170-231\,\si{MHz}$ from the GLEAM survey \citep{Hurley17, For18} and images of \ion{H}{i} emission at $21\,\si{cm}$ separated into \hbox{D-,} \hbox{I-,} and L-components from ATCA and Parkes observations \citep{Kim03, Fukui17, Tsuge24}. Additionally, we incorporated \textit{Herschel} and \textit{Spitzer} infrared imaging at $500$, $160$, $24$, $8.0$, $4.5$, and $3.6\,\si{\mu m}$ from the HERITAGE and SAGE surveys \citep{Meixner10, Meixner13, Meixner06}, and continuum-subtracted emission-line mosaics covering the H$\alpha$, [\ion{S}{ii}], and [\ion{O}{iii}] lines derived from the MCELS survey \citep{Smith99}. 
Special treatment was required by the radio continuum bands, where we smoothed all frequency bands to a common beam size of $5.4\arcmin$, defined by the resolution at the lowest frequency, in order to avoid artifacts for point sources. 
Furthermore, for the optical line emission images, in order to remove the influence of bright point-like sources and of bright-pixel artifacts, we applied a median filter with a radius of $10\arcsec$, resulting in images exclusively containing diffuse structures.

The resulting multiwavelength view of the LMC is displayed in Fig.~\ref{MultImage}. A vast diversity in morphologies is observed: the radio continuum exhibits many discrete sources, likely background active galactic nuclei, as well as \ion{H}{ii} regions such as 30 Dor, emitting thermally via free-free emission \citep{For18}. Importantly, a diffuse glow pervades the galaxy, likely reflecting the distribution of low-energy cosmic rays \citep{Hassani22}, with energies on the order of a few $\si{GeV}$. 
In contrast, as discussed by \citet{Tsuge24}, the \ion{H}{I} D-component traces cold gas filaments and cavities in the LMC disk, whereas the blueshifted I- and L-components are concentrated at the south-east rim. 
A similar network of filaments and cavities is traced by thermal continuum emission from dust, which dominates the emission from $500\,\si{\mu m}$ down to $8\,\si{\mu m}$, corresponding to dust temperatures from $10\,\si{K}$ to several $100\,\si{K}$, with the highest temperatures observed around \ion{H}{ii} regions such as 30 Dor and N11 \citep{Meixner13}. 
At wavelengths below $4.5\,\si{\mu m}$, the continuum emission becomes dominated by main-sequence stars \citep{Meixner06}, which exhibit a distinctly different distribution with the highest concentration of stars in the LMC bar. 
In contrast, the optical emission line mosaics reveal a complex network of interstellar filaments and extended sources, such as SNRs, SGSs \citep{Meaburn80}, and \ion{H}{II} regions, largely tracing regions of massive star formation.  
Finally, the diffuse X-ray emission appears relatively smoothly distributed compared to the much more filamentary structure observed at lower energies. It is particularly interesting to observe the X-ray emission appearing to fill multiple cavities visible in the cooler ISM phases, such as inside the SGSs LMC 4 (labelled H in Fig.~\ref{LabelImage})and LMC 5 in the north, and LMC 2 (labelled B) east of 30 Dor \citep{Sasaki02, Meaburn80}.

\section{Discussion \label{Discussion}}

\subsection{Physical properties of thermal X-ray emission} \label{PhysicalNature}
The eROSITA data set analyzed here provides the deepest view of the LMC as a whole in X-rays, at good spatial and spectral resolution. Using the results of our spatially resolved spectral analysis, we characterize the global properties of thermal X-ray emission (Sect.~\ref{Pressure}). Further, we investigate the impact of modelling assumptions on our results, such as the chosen temperature distribution (Sect.~\ref{MultiComp}), the assumption of CIE (Sect.~\ref{NEITests}), and the neglect of charge exchange emission (Sect.~\ref{CXTest}).

\subsubsection{Mass, pressure, and energetics}\label{Pressure}
Previous studies \citep{Sasaki02, Townsley06, Gulick21} have attributed the hottest emitting gas in the LMC exclusively to the region of 30 Dor. While our study confirms the fact that the highest mean X-ray emitting temperatures can be found around this region (Fig.~\ref{SpecImage}), a hotter and a cooler plasma component appear to be detected across the majority of the galaxy, so that it seems unlikely for these two components to be spatially disjoint \citep[as suggested by][]{Gulick21}. 
Our average densities derived for the hot ISM phase (Sect.~\ref{Spectroscopy}) agree reasonably well with the \textit{XMM-Newton} study of the southeast of the LMC by \citet{Knies21}, who determined densities around $10^{-2}\,\si{cm^{-3}}$ in the southeast. In contrast, \citet{Gulick21} estimated systematically higher densities for the X-ray emitting phase, likely due to the inclusion of bright SNRs in their estimates (see Sect.~\ref{ParMaps}).
Using our derived densities of the hot ISM phase, we can estimate its approximate total mass, by integrating over the entire area of observed X-ray emission, obtaining
\begin{equation}
    M_{\rm hot} = 6.3 \, \left ( \frac{D_{\rm LoS} f}{3\,\si{kpc}} \right )^{1/2}\times10^6\,\si{M_{\odot}}.
\end{equation}
This value, derived over a solid angle of around $20.2\,\si{deg^2}$, implies an average surface density of hot gas of $4.1\times10^{-1}\,\si{M_{\odot}.pc^{-2}}$, which corresponds to a fraction around $3\times10^{-3}$ of the average density of the LMC \citep{Russell92}. 
While the formal statistical errors of these estimates are in the sub-percent range, large systematic uncertainties affect our constraints. These originate from the choice of the physical model for the thermal emission, our criteria for masking compact structures in the emission (associated to, e.g., SNRs), and our chosen boundary of the LMC. 

By combining our constraints on plasma temperatures and densities, we can derive the thermal pressure provided by the hot ISM phase as: 
\begin{equation}
 {P}_{\rm tot} = 10^7 \left( 1 + c_{e}\right) \left( \frac{4\pi \,\left( \eta_1 kT_1^2 + \eta_2 kT_2^2 \right)}{D_{\rm LoS} \,f\,c_{e}} \right )^{1/2},
\end{equation}
which includes the contribution of both ions and electrons. For an ideal nonrelativistic gas, this can be converted into an energy density following $\epsilon = E/V = 3/2\,P_{\rm tot}$. 
The derived distribution of thermal pressure of the hot ISM from our spectral fits and the corresponding energy density, both assuming $D_{\rm LoS} \,f = 3\,\si{kpc}$, is illustrated in Fig.~\ref{EnergyPressure}. 
The highest pressure, with peak values of $P/k \approx 1\times10^5\,\si{K.cm^{-3}}$ is observed in the southeast, in the X-ray spur and around 30 Dor, with a secondary peak coinciding with older stellar populations in the bar. 
A possible explanation for this high inferred pressure in the bar is the contribution of a large number of low-mass stars, which may act as faint unresolved X-ray sources mimicking diffuse emission \citep[e.g.,][]{Wulf19}. 
On the other hand, the pronounced peak in X-ray emission, and equivalently pressure, very close to the southeast rim, while well-known \citep[e.g.,][]{Sasaki02, Wang91}, is a quite peculiar feature of the LMC ISM. While pressures around $10^5\,\si{K.cm^{-3}}$ in 30 Dor are expected due to energy input through massive star formation \citep[e.g.,][]{Malhotra01}, this mechanism appears unlikely to be able to explain the large thermal pressure further south, in the X-ray spur, due to lack of massive stars there \citep{Knies21}. 
On larger scales, the pressure appears to smoothly decrease by up to an order of magnitude, with the lowest values ($P/k \approx 1\times10^4\,\si{K.cm^{-3}}$) apparent at the southern and northern rims. Nonetheless, this value is still comparable to pressures observed in the Galactic ISM, for instance in the local hot bubble \citep{Yeung24}.
On the other hand, the highest pressures we observe here correspond nicely to the lowest hot-gas pressures inferred for compact \ion{H}{ii} regions, which typically occupy the range $10^5-10^6\,\si{K.cm^{-3}}$ \citep{Lopez14}.

By integrating our estimates for the hot-gas energy density over the entire LMC, we estimated the total thermal energy stored in the hot ISM phase, obtaining 
\begin{equation}
    E_{\rm hot} = 8.8 \, \left ( \frac{D_{\rm LoS} f}{3\,\si{kpc}} \right )^{1/2}\times10^{54}\,\si{erg}.
\end{equation}
This value is of a similar order of magnitude to previous estimates  of around $3\times10^{54}\,\si{erg}$ \citep{Gulick21} and $7\times10^{54}\,\si{erg}$ \citep{Wang91}, despite fundamentally different assumed geometries and contamination of the diffuse emission through compact sources. 
Assuming that massive stars provide the dominant energy input into the hot ISM phase, we can estimate its characteristic heating timescale: the total energy corresponds approximately to the energy released by $8800$ supernovae with a ``canonical'' energy of $10^{51}\,\si{erg}$. Assuming an approximate LMC supernova rate of $5\times10^{-3}\,\si{yr^{-1}}$ \citep{Bozzetto17}, 
we can estimate an average rate of energy input $\dot{E} \approx 1.6\times10^{41}\,\si{erg.s^{-1}}$. 
This implies a typical characteristic heating timescale of $t_{\rm heat} \sim 1.8\,\si{Myr}$. A more conservative estimate, corresponding to one supernova per $100\,\si{M_{\odot}}$ formed, would be a current rate of $2\times10^{-3}\,\si{yr^{-1}}$ \citep{Maoz10}, yielding $t_{\rm heat} \sim 4.4\,\si{Myr}$.

Complementarily, using the observed level of X-ray emission, we can compute an approximate cooling timescale, on which radiative energy loss occurs. To achieve this, we assumed cooling in the X-ray band $> 0.1\,\si{keV}$ and a representative metallicity of half solar, and determined the cooling function $\Lambda (T)$ \citep{Sutherland93, Schure09} by integrating the expected X-ray emission over all energies at the temperature measured in each region. The radiative cooling timescale is then given by the ratio of plasma energy density $\epsilon$ and volumetric rate of emission, calculated from the densities and temperatures of the two components:
\begin{equation}
    t_{\rm cool} = \frac{\epsilon}{n_1^2\,\Lambda(T_1) + n_2^2\,\Lambda(T_2)}.
\end{equation}
The resulting distribution of the cooling timescale is shown in Fig.~\ref{EnergyPressure}. While a strong gradient across the galaxy is present, even the smallest observed values are $t_{\rm cool} \sim 100\,\si{Myr}$, more than an order of magnitude larger than the expected heating rate, with the largest values even reaching $t_{\rm cool} \gtrsim1\,\si{Gyr}$. 
Assuming that the energy content of the hot ISM phase is approximately constant over time and not continuously increasing, this discrepancy which is well known on scales from star clusters to galaxies \citep[e.g.,][]{Wang91, Rosen14}, demonstrates the necessity for an additional mechanism for removing the energy stored in the hot ISM phase. 
One possible reason for this mismatch could be enhanced clumping in the hot phase, as the typical cooling time is expected to decrease for a decreasing filling factor as $t_{\rm cool}\propto f^{1/2}$. In addition, efficient optical and ultraviolet radiative cooling in regions reaching temperatures $\lesssim 10^5 \,\si{K}$ may provide an important energy outlet, if such low temperatures can be reached fast in dense clumps \citep{Falle81}, through efficient heat conduction at interfaces with dense shells \citep[e.g.,][]{Rosen14, Steinwandel20}, or through turbulent mixing with colder gas \citep{Lancaster21a, Lancaster21b}. 
Alternatively (or complementarily), an important channel of energy loss could be given by adiabatic expansion and outflows from the ISM into the circumgalactic medium, being driven by overpressure in the hot ISM phase \citep[``galactic fountains'';][]{Shapiro76}. This mechanism may provide an important avenue of feedback on star formation, in particular in low-mass, star-forming galaxies such as the LMC \citep[e.g.,][]{Girichidis16, Nelson19, Hopkins12}.   

\begin{figure*}[t!]
\centering
\includegraphics[width=18.4cm]{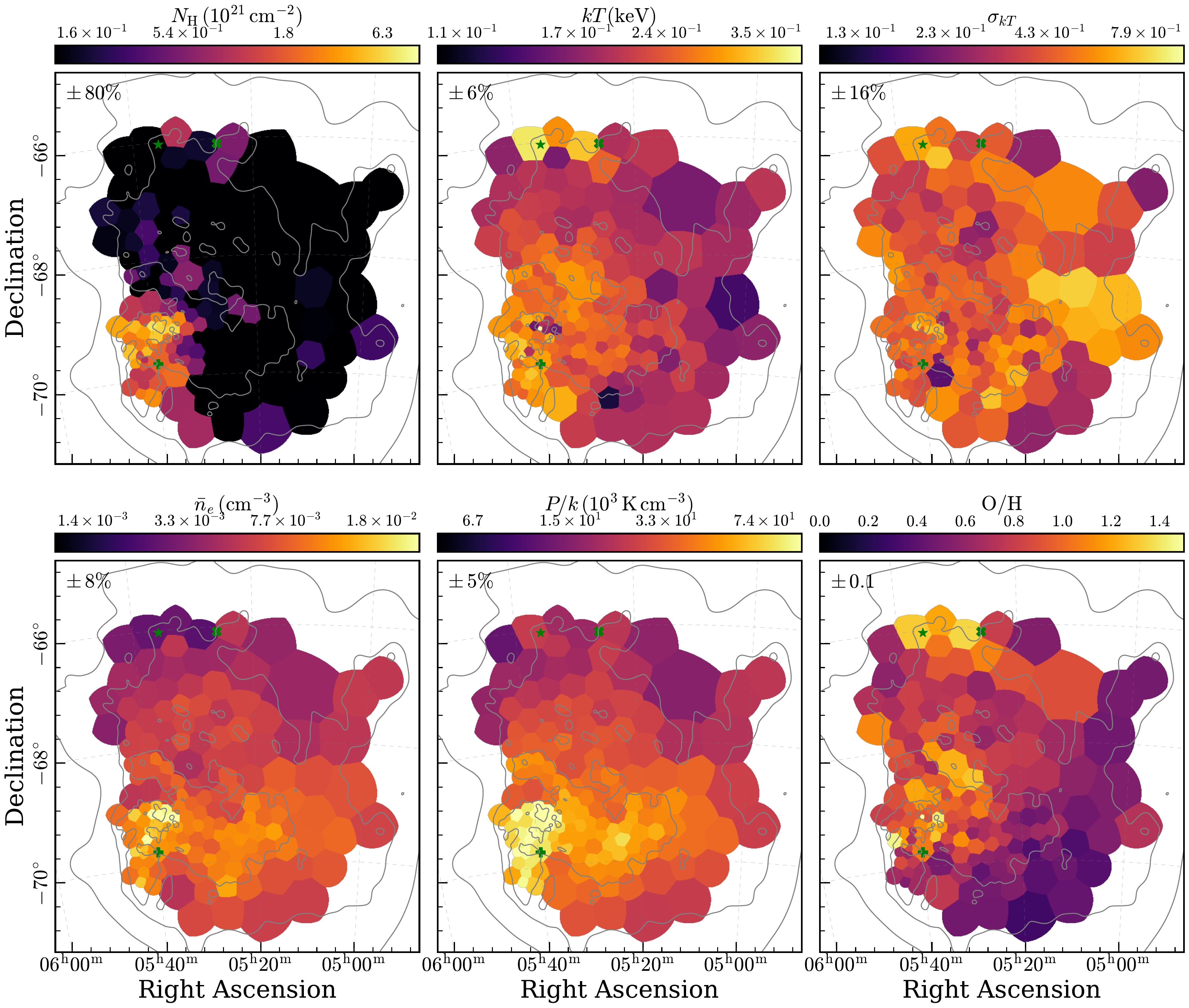} 
\caption{Parameter maps from spectral fitting, assuming a log-normal temperature distribution. The temperature $kT$ corresponds to an emission-weighted mean temperature, whereas $\sigma_{kT}$ describes the characteristic width of the temperature distribution in natural logarithmic scale. All other parameters are identical to Fig.~\ref{SpecImage}.} 
\label{GTNTMaps}
\end{figure*}

\begin{figure}
\centering
\includegraphics[width=\linewidth]{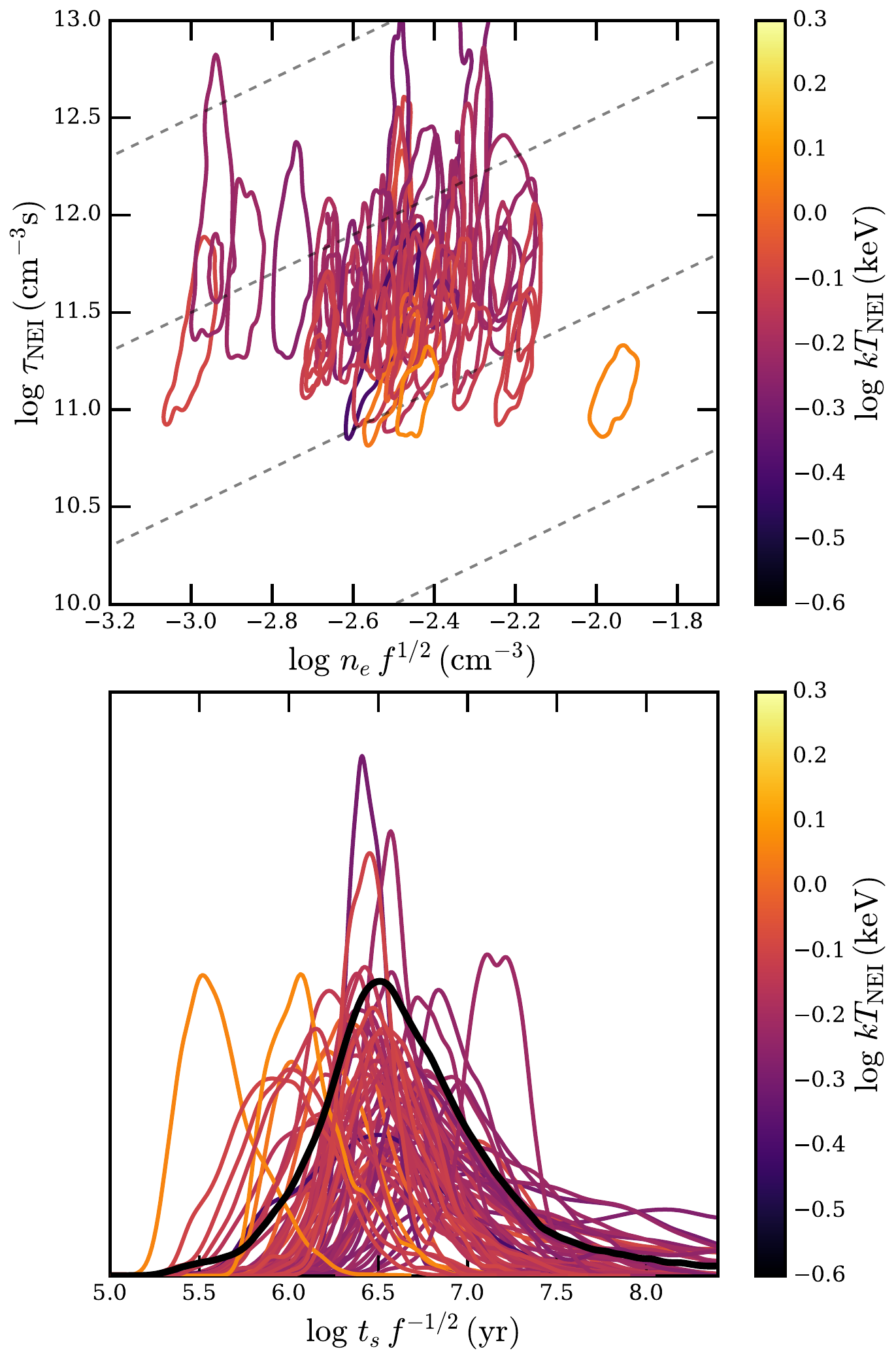} 
\caption{Constraints on NEI in the hot plasma component. The top panel shows the joint $1\sigma$-contours of the density $n_{e}$ and ionization timescale $\tau_{\rm NEI}$ inferred for the hot plasma component in each region. The dashed gray lines correspond to constant shock age $t_{s}$, and are logarithmically spaced between $10^5$ and $10^8\,\si{yr}$.
In the bottom panel, the constraints are converted into probability distributions of $t_{s}$. The thick black line indicates the combined distribution of all regions. 
In both panels, the line colors indicate the electron temperature $kT_{\rm NEI}$ fitted for the respective region. 
} 
\label{NEIProperties}
\end{figure}

\subsubsection{Temperature distribution}\label{MultiComp}
It is a natural question to ask why the ideal model to describe the X-ray emission of the LMC ISM would consist of exactly two plasma components. One might ascribe them to physically distinct origins, such as a diffuse background of hot thin gas, combined with localized regions of cooler, denser gas. 
However, a worthwhile alternative to explore is the presence of a continuous distribution of temperatures, governed by local pressure equilibrium. 
To test this scenario, we used a model of a continuous log-normal temperature distribution \citep[\texttt{vlntd};][]{Cheng21}, rather than the model with two discrete single-temperature plasma components. By repeating our spectral fits across the LMC with this model, we evaluated the dependence of different physical parameters on the assumed model. 

An overview of the resulting parameters is given in Fig.~\ref{GTNTMaps}. We found that the typical fit quality was similar to the original model, implying that neither model is preferred from a statistical point of view.  
However, while qualitative features of the parameter maps in Fig.~\ref{SpecImage} are clearly preserved, quantitative differences in individual parameters are present. 
While the foreground absorption column density is robustly recovered in our alternative model, we observe significantly lower mean temperatures, down to $kT \sim 0.15\,\si{keV}$, particularly in 30 Dor. What persists, however, is the tendency observed in Sect.~\ref{Spectroscopy}, of higher temperatures in the southeast, and colder plasma in the north, west, and southwest. 
Overall, the map of fitted temperature scatter shows little large-scale structure, and exhibits a median value of $\sigma_{kT} = 0.46_{-0.12}^{+0.13}$. For illustration, this value implies that $68\%$ of the emission measure is contributed by plasma with a temperature within a factor $1.6 \pm 0.2$ of the mean temperature. Interestingly, many of the regions with the lowest mean temperature exhibit the highest scatter, implying the necessity for a contribution of hot plasma even in the ``coldest'' regions. 
Regarding the composition of the hot ISM phase, we recovered the significant $\alpha$-element enhancement in the east of the galaxy, with a metal-poor southwest. However, the median oxygen abundance is somewhat increased with respect to the two-temperature model and the typical LMC metallicity, at $\rm O/H = 0.75_{-0.25}^{+0.31}$.   

In order to derive the typical pressure and density in the log-normal model, we followed the derivations outlined by \citet{Cheng21}, taking into account both the mean and the width of the temperature distribution. The maps of pressure and density shown in Fig.~\ref{GTNTMaps} strongly resemble those from the two-component model qualitatively, but quantitative offsets are present. 
A moderate increase can be observed regarding the typical thermal pressure, with values around $20\%$ higher than in Fig.~\ref{EnergyPressure}. Even more strikingly, the median inferred density is $\bar{n}_{e} = 7.6_{-2.5}^{+3.4}\times10^{-3}\,\si{cm^{-3}}$, a relative increase around $40\%$ compared to our original model. 
The reason for both phenomena lies in the implied existence of a low-temperature tail of emitting gas in the continuous temperature model, which is not present in the discrete two-component model. Since, by construction, low temperatures imply large densities due to pressure equilibrium, the inferred average density is increased. 
Unfortunately, given the low but ubiquitous Galactic foreground absorption, our data are only sensitive to the high-temperature tail of the distribution and do not allow us to test for the presence of plasma with $kT < 0.1\,\si{keV}$.
Overall, this experiment of an alternative, but equally justifiable, physical model demonstrates the vulnerability of quantitative spectral properties to systematic errors introduced by modelling assumptions.

\subsubsection{Non-equilibrium ionization}\label{NEITests}
While the choice of plasma models in CIE is commonplace when analyzing diffuse emission of ISM in the LMC \citep[e.g.,][]{Sasaki21, Knies21, Gulick21}, its dynamic nature and the violent processes suspected to energize it motivate the possibility of non-equilibrium ionization (NEI). 
Underionized NEI plasma is frequently observed in SNRs \citep{Borkowski01}, where insufficient time has passed after shock heating for ion-electron temperature equilibration. 
Since SNRs are likely the dominant source of hot plasma in the ISM, we investigated the impact of allowing for NEI in our spectral analysis by replacing one of the CIE components with a plane-parallel shock model \citep[\texttt{vpshock};][]{Borkowski01}. 

The most fundamental result of this effort was that an NEI model did not significantly improve the spectral fit in the vast majority of the LMC, except for a few regions around 30 Dor. This indicates no necessity for recent shock heating of the hot ISM phase, from a statistical standpoint. 
Nonetheless, we can employ our results to place constraints both on the recency of shock heating in the ISM and on the possible effects of NEI on the inferred physical parameters of the LMC.
In Fig.~\ref{NEIProperties}, we indicate our probabilistic constraints on the fitted ionization age (indicating the strength of NEI), in dependence of both the temperature and inferred electron density of the hot component. In all cases, the observed spectra allow at most for weak departure from CIE \citep[see][]{Smith10}, with ionization ages $\tau_{\rm NEI} > 10^{11}\,\si{cm^{-3}.s^{-1}}$. Similarly, the typical temperatures fitted to the hot component are only weakly enhanced with respect to the CIE values, at $kT_{\rm NEI} = 0.60_{-0.08}^{+0.12}\,\si{keV}$.
Similarly, changes on both the inferred typical density and pressure are below $5\%$, indicating that CIE is likely a good assumption in spectral fitting.. 

We can exploit our results to constrain the minimum typical heating timescale of the ISM in the following manner: The ionization age, specifying the degree of equilibration between electrons and ions in the hot plasma \citep{Borkowski01}, parametrizes the number of Coulomb collisions per particle since shock heating. Hence, we can estimate the ``shock age'', the time passed since the assumed initial heating of the ISM, as 
\begin{equation}
    t_{s} = \frac{\tau_{\rm NEI}}{n_{e}^{\rm NEI}},
\end{equation}
where ${n_{e}^{\rm NEI}}$ is the electron density of the hot component. 
The constraints on $t_{s}$ from our individual regions are shown in Fig.~\ref{NEIProperties}, with two properties standing out: first, the peak of the distribution is at around $t_{s} \approx 3\,f^{1/2}\,\si{Myr}$, which one may interpret as the typical timescale of shock heating of the hot ISM phase. Intriguingly, this agrees very well with the typical supernova heating timescale inferred in Sect.~\ref{Pressure}, supporting the picture of massive stars as the main energizer of the hot ISM phase.
Second, a single region stands out from the sample, exhibiting the most recently heated plasma with $t_{s}\sim0.3\,f^{1/2}\,\si{Myr}$ and a high temperature of $kT_{\rm NEI} \sim 1 \,\si{keV}$. This region is located in the center of 30 Dor \citep[see][]{Townsley06, Cheng21}, surrounding the central young cluster of massive stars R136 \citep{Crowther10}. Recent heating in this active region appears highly plausible \citep{Sasaki21}, as we are likely witnessing the combined action of winds from numerous supermassive and Wolf-Rayet stars, rather than  supernovae, due to its low age \citep{Crowther16}.
The true timescale of energy input in 30 Dor may even be significantly smaller than $10^5\,\si{yr}$, as our constraint was calculated assuming a path length of $3\,\si{kpc}$ for the emitting plasma, which is likely a severe overestimation.

\subsubsection{Contribution of charge-exchange emission}\label{CXTest}
Given the likely large-scale mixing of cold and hot ISM phases, a possible additional source of diffuse X-ray emission from the ISM is charge exchange (CX) emission at the interface of ionized and neutral media \citep{Lallement04}. For instance, such CX emission has been reported to likely constitute a large fraction of the thermal X-ray emission in neighboring spiral galaxies, such as M51 \citep{Zhang22}. 

We can parametrize the expected contribution of CX to the X-ray emission in the following manner: the CX ``emission measure'' depends on the product of the densities of cold, neutral ``donor'' material $n_{\rm H}$ and hot ionized ``recipient'' material $n_{i}$, integrated over the emitting volume $V$ \citep{Zhang22}
\begin{equation}
    \mathrm{EM_{CX}} = \frac{1}{4\pi D^2}\int \mathrm{d}V n_{\rm H}n_{i}, 
\end{equation}
where $D$ is the distance to the source. Given the large CX cross-section, regions of neutral material are likely sufficiently dense to prevent an ion from escaping before being neutralized. Hence, we can assume an emitting volume with a characteristic depth equivalent to the mean free path of the ions under CX, $V = A/(\sigma n_{\rm H})$ \citep{Zhang22}. 
Assuming a uniform typical density within the volume, we obtain
\begin{equation}
    \mathrm{EM_{CX}} = \frac{n_{i} A}{4\pi D^2 \sigma}, 
\end{equation}
where $A$ is the area of the emitting interface, and $\sigma \approx 3\times10^{-15}\,\rm cm^2$ is the typical CX cross-section.    
Parametrizing the relation between the interface area $A$ and the geometric surface area of the galaxy as $A = f_{A} \Omega D^2$, we remove the distance dependence:
\begin{equation}
    \frac{\mathrm{EM_{CX}}}{\Omega} = \frac{n_{i} f_{A}}{4\pi \sigma}, 
\end{equation}
where $\Omega$ is the angular size of the galaxy, and $f_{A} \sim \mathcal{O}(1)$ is a dimensionless factor governed by the turbulent mixing between the hot and cold ISM phases. 

In order to estimate the expected flux level of CX emission, we used the \texttt{acx2} model\footnote{See \url{https://github.com/AtomDB/ACX2} for a detailed description of the model.} \citep{Smith12} which predicts the following X-ray flux in the $0.2-2.3\,\rm keV$ band for a plasma temperature of $0.25 \,\rm keV$, elemental abundances of half solar, and a characteristic ion velocity of $v = 100\,\rm km/s$:\footnote{For comparison, the typical thermal speeds $v = \sqrt{2\,kT/m}$ would be around $v \approx 220\,\si{km.^{-1}}$ for a proton and $v \approx 55\,\si{km.^{-1}}$ for an oxygen ion.}
\begin{equation}
F_{\rm CX} = 4.2\times10^{-20} \, \left( \frac{v}{100 \, \rm km\, s^{-1}}\right)\,\mathrm{EM_{CX}}\,\mathrm{erg\,cm^{-2}\,s^{-1}}.   
\end{equation}
Finally, we obtain the following prediction of the surface brightness $\Sigma_{\rm CX} = F_{\rm CX}/\Omega$ of charge exchange expected for the LMC, dependent on $f_{A}$, $n_{i}$, and $v$:
\begin{eqnarray}
    \Sigma_{\rm CX} &=& 4.8\times10^{-16} \,\mathrm{erg\,cm^{-2}\,arcmin^{-2}\,s^{-1}} \\
    &\times& f_{A} \, \left( \frac{v}{100 \, \rm km\, s^{-1}}\right)\,\left( \frac{n_{i}}{5\times10^{-3}\,\rm cm^{-3}}\right),
\end{eqnarray}
which corresponds to around $7\%$ of the average surface brightness of diffuse thermal emission in the LMC, \mbox{$\Sigma_{\rm LMC} = 7.3\times10^{-15}\,\mathrm{erg\,cm^{-2}\,arcmin^{-2}\,s^{-1}}$}. 
Hence, whether or not the effect of charge exchange is negligible in spectral analysis depends strongly on the density and geometry of a given emitting region. Rarefied regions with turbulent mixing are likely to exhibit a strong relative CX contribution, as the ratio of fluxes of a CIE plasma and a CX component scales as $\Sigma_{\rm CIE}/\Sigma_{\rm CX} \propto n_{i}\,f_{A}^{-1}$.    
\begin{figure}[t!]
\centering
\includegraphics[width=0.8\linewidth]{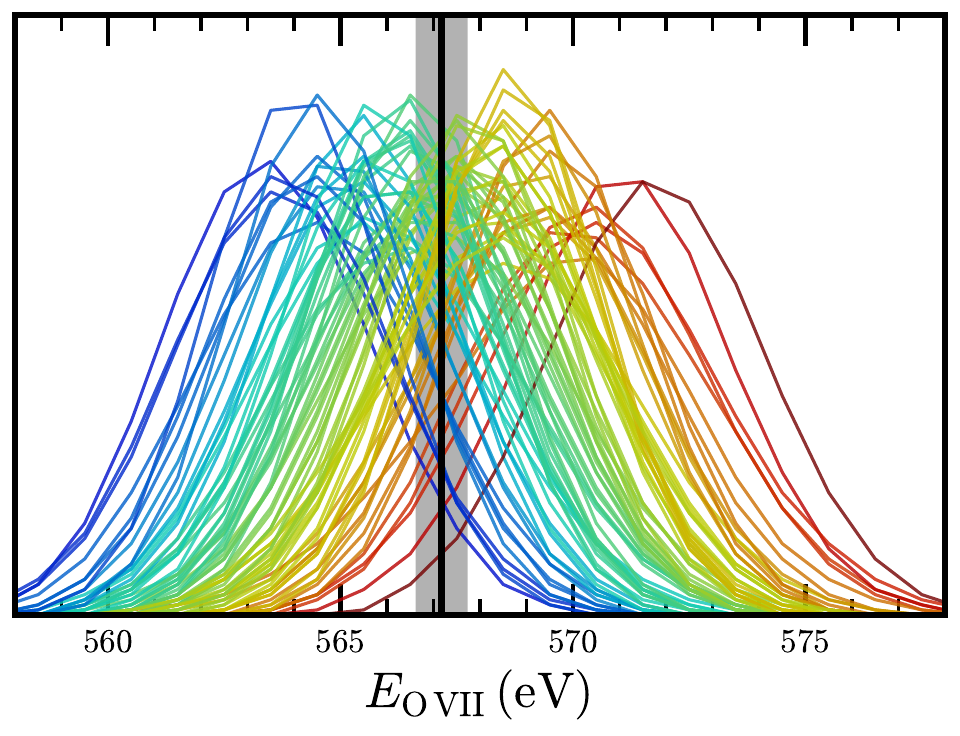} 
\includegraphics[width=0.8\linewidth]{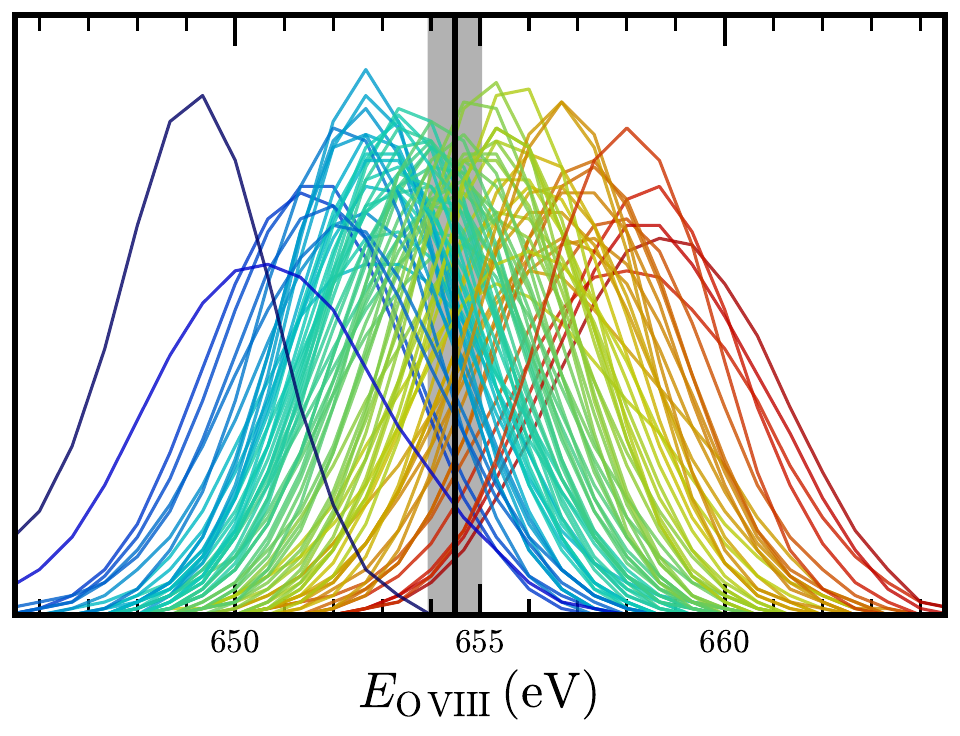} 
\caption{Centroid determination of \ion{O}{vii} (top) and \ion{O}{viii} (bottom) lines. In both panels, we display the probability distribution functions for the centroid energy of all regions where both lines were detected with at least $10\sigma$ significance.  
The vertical black lines mark the mean line centroid energies, constructed via a weighted average over all regions, and the shaded areas indicate the corresponding $3\sigma$ uncertainties. The color scale reflects the mean centroid energies, purely for the purpose of visual differentiation.  
} 
\label{LineEnergies}
\end{figure}

Our eROSITA X-ray data lack the spectral resolution to unambiguously identify CX features, for instance by resolving individual lines in the He$\alpha$ triplet of \ion{O}{vii} \citep{Zhang14, Zhang22}. However, we can attempt to look for such features indirectly, by determining the effective centroid energy of the \ion{O}{vii} line complex:
In order to achieve this, we repeated our spectral fitting effort, excluding the contribution of oxygen from the thermal emission by setting $\rm O/H = 0$, and replacing it with two Gaussian lines at the location of the \ion{O}{vii} triplet around $570\,\si{eV}$, and of the \ion{O}{viii} Ly$\alpha$ line around $654\,\si{eV}$. While the absolute energy calibration of eROSITA is still uncertain on the order of a few $\si{eV}$ \citep{Merloni23}, this allows us to calibrate the centroid energy of \ion{O}{vii}, which is sensitive to CX, using the relative offset to \ion{O}{viii}.  
The line width was fixed to $10\,\si{eV}$, much below the spectral resolution \citep{Predehl21}, but above the energy spacing of the response matrix, and line positions and amplitudes were left free to vary. 

The resulting constraints on centroid energies are displayed in Fig.~\ref{LineEnergies}. While the typical errors of individual centroid energies are around $2\,\si{eV}$, computing the weighted average over all regions with $>10\sigma$ significance yielded precise values, $E_{\ion{O}{vii}} = 567.42 \pm 0.18 \,\si{eV}$ and $E_{\ion{O}{viii}} = 654.56 \pm 0.18 \,\si{eV}$. 
For line emission from a CIE plasma with a temperature around $0.25\,\si{keV}$, our estimated intrinsic line centroid of the \ion{O}{viii} complex, taking into account the \ion{O}{vii} line at $666\,\si{eV}$, is $E^{\prime}_{\ion{O}{viii}} = 654.89 \,\si{eV}$ \citep{AtomDB}. Hence, our best estimate for the mean  He$\alpha$ centroid energy is $E^{\prime}_{\ion{O}{vii}} = 567.75 \pm 0.25 \,\si{eV}$.  
From this value, we can derive a crude estimate of the \ion{O}{vii} G-ratio, to further characterize the ionization state \citep{Mewe78, Porquet01}. It is defined as $G = (f+i)/r$, where $f$, $i$, $r$ are the forbidden, intercombination, and resonance line fluxes of the He$\alpha$ triplet. Following \citet{Zhang22}, we assume $f/i=4.44$, and combine our centroid measurement with the known intrinsic line energies \citep{AtomDB} to obtain $G = 1.15\pm0.10$. While typically $G \lesssim 1.4$ is considered to be consistent with CIE \citep{Porquet01, Zhang22, Wang12}, a nonzero contribution of CX emission is not excluded by our measurement. In fact, a value of $G \gtrsim 1$ is only expected for very cold plasmas \citep[$kT\lesssim0.1\,\si{keV}$;][]{Sun25}, and the observed line centroid could be reconciled with a superposition of CIE and CX emission, with the latter contributing about one quarter of the \ion{O}{vii} flux. 
Nonetheless, while the formal statistical uncertainty of our estimate appears quite convincing, there are clear systematic effects which might impact our measurement. These include the possibility of a non-constant miscalibration of the energy scale, a shift of the measured line centroid through contamination by lines of other species. Furthermore, additional astrophysical effects could in principle modify relative line fluxes, such as enhanced recombination or resonant scattering \citep[see][]{Wang12, Porquet10}.   

\begin{figure*}[t!]
\centering
\includegraphics[width=18.4cm]{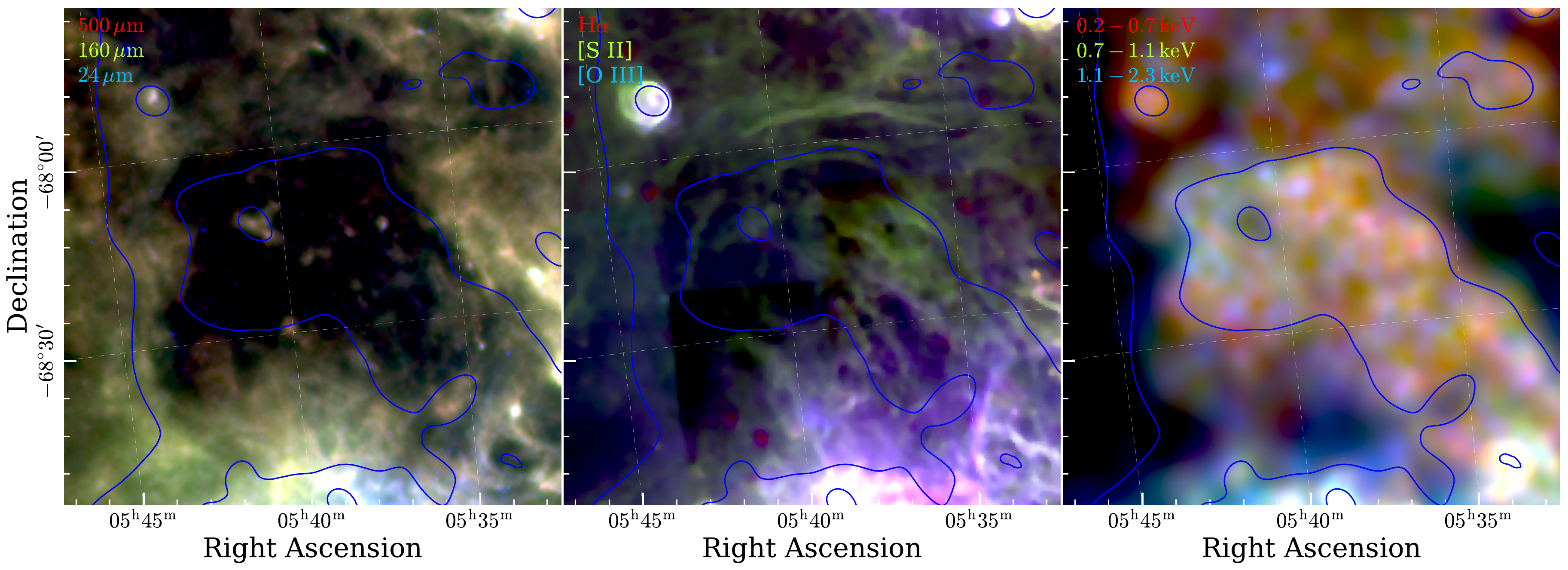} 
\caption{Multiwavelength view of the region of SGS 17 \citep{Kim99}. This figure is as Fig.~\ref{MultImage}, but zoomed in on an $80\arcmin$ box around SGS 17, and showing only far-/mid-infrared, optical emission line, and X-ray images. } 
\label{OutflowImage}
\end{figure*}

From a physical point of view, one of the main benefits of including a CX component in our model is that it elegantly reduces the number of components with separate temperatures, as a single-temperature hot ISM component is capable of emitting thermal bremsstrahlung and line emission, as well as CX emission. Hence, such a model could provide a softer (CX) and harder (\texttt{vapec}) emission component without requiring two physically distinct plasma components. 
We tested this hypothesis by repeating our fits, replacing one \texttt{vapec} with a \texttt{vacx} \citep{ACX} component, with linked temperatures and abundances. 
While this effort returned a realistic typical plasma temperature at $kT \sim 0.25\,\si{keV}$, and mean density at $n_{e} \sim 10^{-2}\,\si{cm^{-3}}$, the statistical quality of the fits was significantly deteriorated compared to our original model.
This is evidenced for instance by model comparisons using an approximate Bayes factor computed from our posterior samples \citep[as in][]{Mayer23}, indicating the relative likelihood of the models with and without CX, which evaluates to $\ln \mathcal{B} = \ln \mathcal{L}_{\rm CX}/\mathcal{L}_{\rm CIE} = -20^{+14}_{-12}$ across all regions. A more traditional metric is given by the reduced $\chi^2$ statistic of the model, which is found to be in the range $\chi^2_{\rm red} = 1.35^{+0.26}_{-0.19}$, significantly worse than our main model (see Sect.~\ref{Spectroscopy}). 
Therefore, while providing an attractive simplification of the physical picture giving rise to the thermal X-ray emission, the scenario of a single-temperature plasma producing additional CX emission is unable to reproduce the observed X-ray spectra satisfactorily.

\subsection{Relation of hot, warm, and cold ISM phases}
The X-ray data and analysis presented in this work complement the rich multiwavelength picture available of the ISM in the LMC, and allow us to study the correlation between the distribution and properties of X-ray emitting hot ISM, and colder gas and dust. While we leave a more quantitative study of the correlation of the different ISM phases and stellar populations to paper II, a few observations stand out clearly when inspecting our maps of X-ray and multiwavelength emission (Figs.~\ref{SpecImage} and \ref{MultImage}):

First, within the boundaries marked by the cold gas of the LMC disk, our map of diffuse X-ray emission reveals a very smoothly distributed hot ISM phase, in particular in the softest X-ray band, at $0.2-0.7\,\si{keV}$. A few apparent gaps and holes in the X-ray emission are visible, which correspond to peaks in the distribution of the cold phase, traced by \ion{H}{I} and far-infrared emission. This includes the X-ray dark region west of the X-ray spur \citep{Knies21}, which, while exhibiting enhanced foreground absorption, is found to have a much lower typical density and lower temperature than in the spur (Fig.~\ref{SpecImage}). 
Conversely, cavities visible in the cold ISM tend to be filled with hot X-ray emitting gas, such as the example region of SGS 17 \citep[region E in Fig.~\ref{LabelImage};][]{   Kim99}, displayed in Fig.~\ref{OutflowImage}. This effect does not appear to depend on the energy band, and the foreground absorption across the majority of the galaxy was constrained to be very low in spectral fitting. Hence, we argue that this is a manifestation of the physical distribution of the different ISM phases, which may indicate a high volume filling factor of the hot phase. 

Second, our map of foreground absorption column density intrinsic to the LMC reveals a strongly absorbed band in the southeast ($N_{\rm H} \gtrsim 10^{21}\,\si{cm^{-2}}$), in very good agreement with previous works \citep{Sasaki21, Knies21}, while the rest of the LMC appears to exhibit no significant absorption. The southeast rim with its high absorption overlaps almost exactly with the spatial morphology of the I- and L-components of \ion{H}{i} emission \citep{Tsuge19, Tsuge24}. These components extend from the approximate location of 30 Dor to the south, far beyond the region of bright X-ray emission. 
This finding supports the notion that there are significant amounts of intervening cold gas located ``in front'' of the LMC, as implied by the relative blueshift of the \ion{H}{i} signal. These gas components connect smoothly to the Magellanic Bridge, a stream of cold gas created through tidal interaction of the Galaxy with the Magellanic Clouds \citep{Mathewson74, Gardiner96}. Based on the observed lack of warm and hot ISM phases or stellar mass coincident with I- and L-component, this intervening component appears to consist of exclusively cold gas south of $\delta \sim -70.5^{\circ}$.

\begin{figure*}[t!]
\centering
\includegraphics[width=18.4cm]{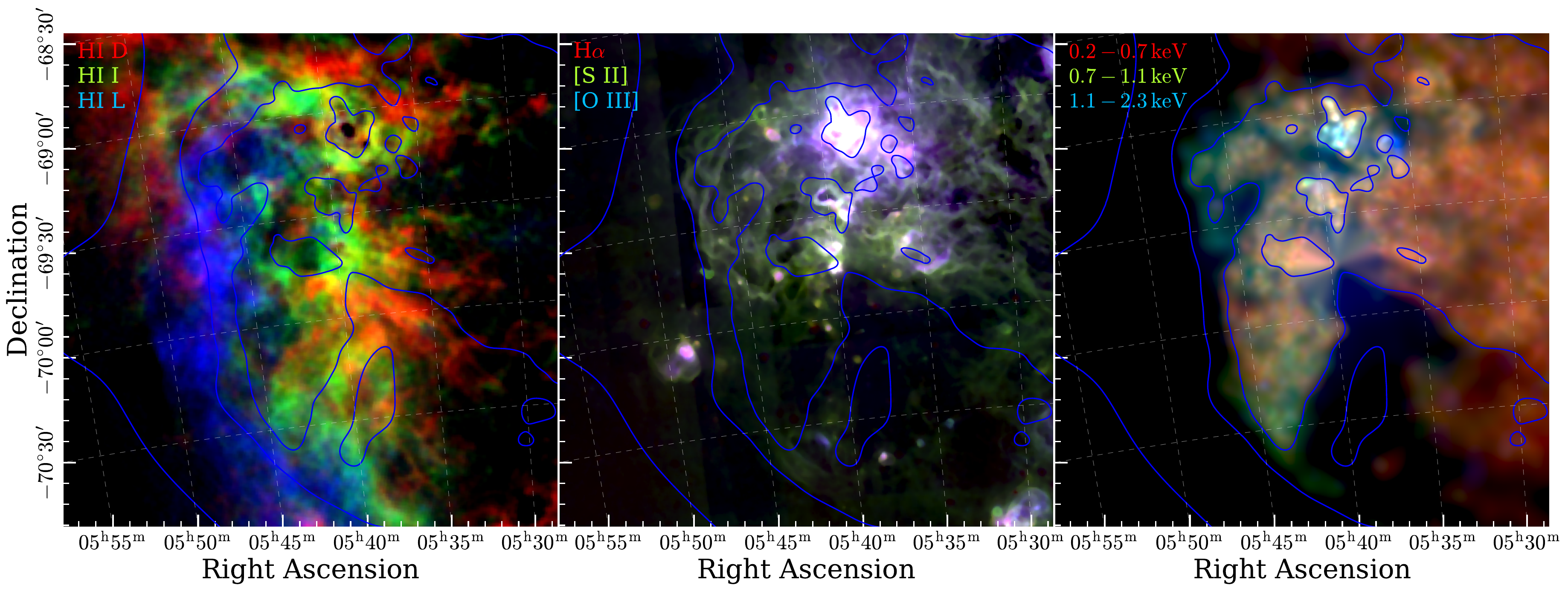} 
\caption{Multiwavelength view of the X-ray spur. This figure is as Fig.~\ref{MultImage}, but zoomed in on a $150\arcmin$ box around the spur, and showing only \ion{H}{i} emission line intensity, optical emission line intensity, and X-ray images. } 
\label{SpurImage}
\end{figure*}

Finally, a region which stands out clearly in X-rays is that of the \ion{H}{ii} region 30 Dor and the X-ray spur, due to its bright and comparatively hard diffuse emission. A zoom-in on this region, displaying the emission of cold, warm, and hot gas, is shown in Fig.~\ref{SpurImage}. 
The complex and intense X-ray emission within 30 Dor agrees well with the intense H$\alpha$ emission typical for \ion{H}{ii} regions. However, the bright X-ray emission further south in the X-ray spur lacks any counterpart in optical line emission, which would indicate the presence of a warm gas component,  
and does not coincide with any population of massive stars. 
Further, the cold gas does not share the morphology of the spur, instead the hot X-ray bright gas seems to be located between the brightest filaments of \ion{H}{i} emission in the I- and L- components \citep[see Fig.~14a in][]{Knies21}. 
These properties of the X-ray spur were studied in great detail by \citet{Knies21}. They proposed a scenario in which the collision of the different cold gas components has led to an enhancement in the emission of the hot-gas component through compression, and  triggered the star formation in 30 Dor, a few $\rm Myr$ ago. This is supported also by the location of the X-ray spur west of the L-component (i.e., trailing behind in the suspected direction of motion), spatially closer to the I-component, which is hypothesized to be the product of said interaction \citep{Tsuge19, Knies21}.

Our spectral analysis (Fig.~\ref{SpecImage}) confirms one of the main findings of \citet{Knies21}: in a two-component thermal plasma model, the peak of the hotter X-ray emitting component accurately reflects the morphology of the X-ray spur, while the colder component is less peaked there. 
This observation appears to indicate physically separate origins of the two X-ray bright plasma components, with the hotter one reflecting plasma heated by localized mechanisms, like colliding gas flows for the X-ray spur. 
However, looking at other metrics in our spectral analysis provides a much less clear picture. While our emission-weighted mean temperature exhibits a broad peak in the southeast of the LMC, it shows very strong scatter, and the X-ray spur does not exhibit enhanced mean temperatures compared to its surroundings.
A similar picture emerges when modelling the emission with a continuous temperature distribution (see Fig.~\ref{GTNTMaps}), in which no notion of a discrete hotter component exists. Instead, the characteristic feature of the X-ray spur, preserved in different spectral models, is that of very large thermal pressure. 
The observation of a localized enhanced pressure in the X-ray spur is naturally compatible with the idea of its formation, involving compression through collision of cold gas components \citep{Knies21}. An additional enhancement of the heating efficiency may be given by the eastward proper motion of the LMC \citep{Kallivayalil13, Gaia18c}, which may enhance collision velocities along the leading (eastern) edge of the galaxy.  

An interesting observation in this context is that of lower $\alpha$-element abundances in the X-ray spur, compared to regions north of it (Figs.~\ref{SpecImage} and \ref{GTNTMaps}). While the observed values are rather typical for the LMC (i.e., $\rm O/H\sim0.5$), this argues for a separate origin of the heated ISM compared to regions with enhanced abundances, which have likely received energy input and $\alpha$-enrichment from young massive stars. Clearly, the scenario of compression-induced heating does not imply any enhancement with $\alpha$-elements typical for massive stars, and is consistent with this observation. 
Alternatively, we consider the possibility that the inflow of metal-poor gas from the SMC, driven by tidal interactions \citep{Fujimoto90}, may have contributed to the suppression of $\alpha$-element abundance in the spur. 
Such a tidally induced inflow is both suggested theoretically \citep{Bekki07}, and supported observationally, given the low dust-to-gas ratio in the \ion{H}{i} ridge, traced by its weak sub-mm emission observed with \textit{Planck} \citep{Fukui17}. Additionally, a metal-poor stellar population appears to be the present in the LMC, pointing towards past accretion from the SMC \citep{Olsen11}.

\subsection{The composition of the ISM}
\subsubsection{Abundances and enrichment through massive stars} 
Our spectral analysis provides us with the first characterization of the distribution of different elements in the hot ISM phase, both in a relative and an absolute sense. While the quantitative abundances were somewhat dependent on the model we used, we can note definite deviations from the typical LMC abundances (about half solar) in oxygen, neon, and magnesium in Figs.~\ref{SpecImage} and \ref{GTNTMaps}. As iron was fixed to $\rm Fe/H = 0.5$, this implies deviations from the solar composition also in a relative sense, as $\rm \alpha/Fe > 1$ in many regions. This observation is a strong indicator of the chemical enrichment of the hot ISM in the LMC by young massive stars, the main producers of oxygen, neon, and magnesium \citep{Sukhbold16}. Since the typical ISM heating timescale appears to be rather short ($\sim10^6\,\si{yr}$; Sect.~\ref{Pressure}), this finding likely traces recent star formation \citep[see also][]{Sasaki21}. 
The fact that the region of enhanced $\alpha$-abundances appears to broadly coincide with the location of the star-forming region 30 Dor, rich in very massive stars \citep{Kennicutt88}, agrees very well with this idea.
In addition, we would like to point out, that, while the relative distributions of the different light elements agree very well with each other, neon appears to be globally overabundant with respect to oxygen, with a typical ratio of $\rm Ne/O \sim 1.5$. If physical, this finding is not naturally expected in the scenario of massive-star-driven ISM enrichment, which would be expected to approximately reproduce the solar abundance ratio \citep{Sukhbold16}. However, it is also plausible that such an apparent overabundance is spuriously produced by an inappropriate spectral model, as the relative intensity of oxygen and neon emission lines depends strongly on the chosen model and fitted temperature.  

A possible point of concern lies in the strong correlation of the distribution of relative elemental abundances (e.g., $\rm O/H$) with the mean plasma temperature in Fig.~\ref{SpecImage}. Both maps feature the highest values, indicating strong $\alpha$-enrichment and hot plasma, respectively, in an extended region in the southeast of the LMC, with lower values prevailing in the southwest and west of the galaxy. Again, a possible explanation could be a poorly chosen spectral model, inducing spuriously higher $\alpha/\mathrm{Fe}$ ratios for higher temperatures. 
However, the correlation between the parameters is not perfect, showing significant systematic deviations. Hence, even if one were to assume that the entirety of the correlation between $kT_{\rm mean}$ and $\rm O/H$ were spurious, one would still find a strong global asymmetry across the LMC, with the north and east exhibiting higher $\alpha$-element abundances for a given temperature than the south and west (including the stellar bar). A similar picture also emerges in our model with a continuous plasma temperature distribution (Fig.~\ref{GTNTMaps}). 
Such a macroscopic difference appears quite sensible given the concentration of low-mass stars in the bar, in contrast with the existence of many associations of young massive stars in the north and east (Fig.~\ref{MultImage}). 

\begin{figure}[t!]
\centering
\includegraphics[width=\linewidth]{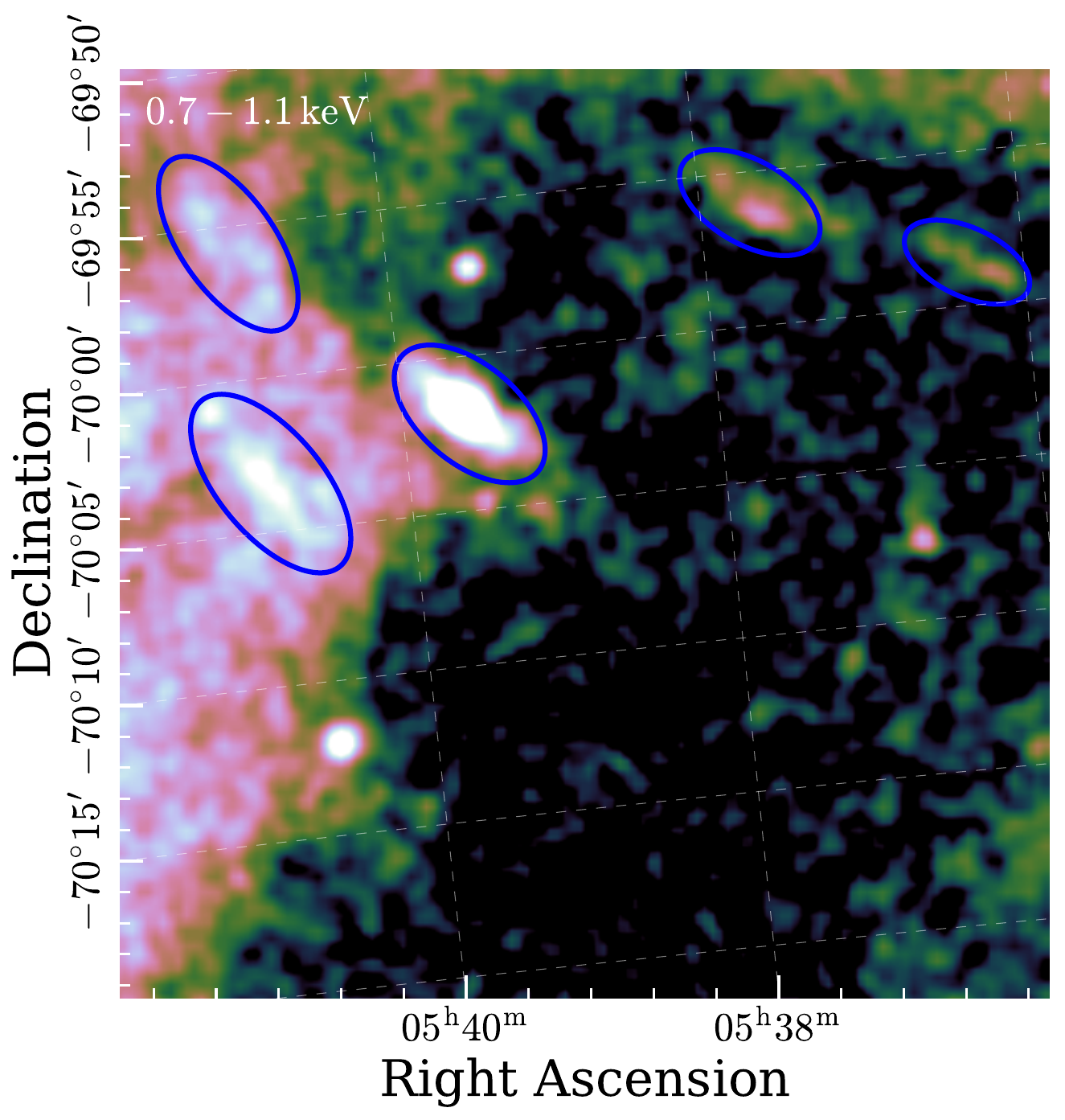} 
\caption{Filaments in a $30\arcmin$-diameter box south of LMC X-1 in the intermediate X-ray band $0.7-1.1\,\si{keV}$. We show an exposure-corrected image, smoothed with a $15\arcsec$ Gaussian kernel. 
The blue ellipses highlight the approximate size and orientation of the identified structures.
} 
\label{Filaments}
\end{figure}

\subsubsection{Potential signatures of outflows}
Given the extensive overabundance in oxygen, neon, and magnesium we detect through spectral analysis, large-scale enrichment from massive stars appears likely. In this context, it is intriguing to inspect Fig.~\ref{OutflowImage}, which displays the multiwavelength view of a shell labelled SGS 17 by \citet{Kim99}, centered on \mbox{$\alpha, \delta = 05^{\rm h}40^{\rm m}30^{\rm s}, -68^{\circ}19^{\prime}$}, around $1^{\circ}$ north of 30 Dor. 
Clearly, we see a near-perfect anticorrelation between the distribution of hot gas, traced by diffuse X-rays, and the shell-like structure visible in the emission of cold dust in the infrared. Intriguingly, optical emission lines, in particular [\ion{S}{ii}],  exhibit a highly filamentary structure, both on the northern rim of the shell \citep{Book08}, and on its inside. 
Given its location in a metal-rich region, one may interpret these emission-line filaments as signatures of metal-rich outflows into the cavity. Given the brightness of optical line emission on the outskirts of the cavity, it is conceivable that the hot gas filling the shell originated in massive star-forming regions embedded in the dense gas, most prominently 30 Dor.

\subsection{``Green'' filaments}
When inspecting the unfiltered multi-band image of the X-ray emission of the LMC (Fig.~\ref{RGBImage}), we noticed a group of interesting elongated features, emitting mostly in the intermediate energy band $0.7-1.1\,\si{keV}$, located south of LMC X-1. A zoomed-in view of these features, which are also apparent in the \textit{XMM-Newton} data of the region \citep[e.g., Fig.~4 of][]{Knies21}, is displayed in Fig.~\ref{Filaments}. 
The brightest one of these features is considered an SNR candidate, but the lack of any associated extended multiwavelength emission makes this interpretation uncertain \citep{Zangrandi24}. 
The filaments displayed in Fig.~\ref{Filaments} share several characteristics with each other, indicating a possible common physical origin: they are all dominated by emission in the $0.7-1.1\,\si{keV}$ band, which may be indicative of neon or iron line emission, or an unusually peaked continuum. They all appear strongly elongated, perhaps even filamentary, and are oriented roughly from northeast to southwest. Finally, none of them appear to have a significant multiwavelength counterpart at optical, infrared, or radio energies.

While we are highly uncertain about the physical nature of these filaments, one may speculate that they could be a further manifestation of a metal-rich outflow from the nearby massive young stars. However, this scenario may struggle with explaining their peculiar shape and isolation in individual clumps. One scenario is provided by simulations of colliding shells in the ISM, with density contrasts leading to ablation of the clumps \citep[e.g.,][]{Krause13}, creating similar morphologies to the observed ones.
Alternatively, since the filaments are located in a strongly absorbed region, one could assume that their apparent peculiar shape is simply an artifact of a smoothly distributed hot X-ray emitting plasma, combined with localized ``holes'' in foreground absorption. However, neither \ion{H}{I} nor far-infrared emission exhibit any such holes expected for a local lack of cold absorbing material. 
Finally, one may also suspect the scattering of radiation from a bright nearby X-ray source, for instance LMC X-1, to play a role. However, also here, it is unclear why the morphology would be highly filamentary, contrasting typical scattering halos \citep{Predehl95}, and why the suspected dust scattering screen would be invisible in the infrared.
In any case, the quasi-alignment and elongated shape of the observed filaments may imply that their common orientation is dictated by magnetic fields. In fact, magnetic fields are known to play an important role in the region south of 30 Dor, as evidenced by strongly polarized filamentary radio emission  in the vicinity \citep{Mao12}.

\subsection{Nonthermal Emission: 30 Dor and SGS \mbox{LMC 2}} \label{NonthermalStuff} 

As a side product of our study of hot-gas properties, our spectral analysis allowed us to constrain the brightness of nonthermal X-ray emission across the LMC.  
As shown in Fig.~\ref{SpecImage}, there is virtually no detected nonthermal emission in regions outside the southeast. In those regions, the surface brightness of nonthermal X-ray emission can typically be constrained at $\Sigma_{\Gamma} < 5\times10^{-16}\,\si{erg.s^{-1}.cm^{-2}.arcmin^{-2}}$ at $95\%$ confidence in the $1.0-5.0\,\si{keV}$ band. 
Instead, the map of ``nonthermal'' emission is clearly dominated by the surroundings of LMC X-1 \citep{Mark69}, whose dominant emission is a thermal continuum from its accretion disk, exhibiting a typical temperature of $\sim 1\,\si{keV}$ \citep{Nowak01}. However, due to its strong intrinsic absorption and high temperature, its emission may mimic that of a hard powerlaw in the most sensitive range in our spectral fits $\sim 1.0-2.3\,\si{keV}$, which explains our hard fitted value of $\Gamma=0.90\pm0.18$ in the region closest to the object. 

In spite of this, we argue that, in several regions surrounding the \ion{H}{ii} region 30 Dor, our detected nonthermal component is in fact reflecting physical synchrotron emission. This includes the center of 30 Dor, where strong nonthermal synchrotron emission originates from the massive young star cluster R136, aged $1.5\,\si{Myr}$ \citep{Crowther16}. A potential additional contribution is given by older stars up to $5\,\si{Myr}$ \citep{Schneider18}, which may have already exploded, leaving behind SNRs. 
Our fit of the power-law component in 30 Dor yields a spectral index of $\Gamma = 2.02\pm0.35$, and an average surface brightness $\Sigma_{\Gamma} = (2.3\pm0.3)\times10^{-14}\,\si{erg.s^{-1}.cm^{-2}.arcmin^{-2}}$, corresponding to a total nonthermal flux of $F_{\Gamma} = (5.7\pm0.8)\times10^{-13}\,\si{erg.s^{-1}.cm^{-2}}$ in the $1.0-5.0\,\si{keV}$ band. While our analysis was of course not aimed at studying well-defined compact emission regions, our photon index agrees within rather large errors with the constraints on R136 based on an early eROSITA observation by \citet{Sasaki21}. However, there is a discrepancy with other measurements yielding somewhat steeper X-ray emission, \citep[e.g., $\Gamma=2.56\pm0.06$;][]{Cheng21}, possibly caused by our exclusion of point sources. 
A further region of significant synchrotron emission is the well-known superbubble 30 Dor C \citep{Dennerl01} around $15\arcmin$ to the west of R136, likely inflated by the young OB association LH 90 \citep{Testor93}. This shell is estimated to be $\sim4\,\si{Myr}$ old \citep{Smith04} and is one of the few superbubbles visible in TeV $\gamma$-rays \citep{HESS15b} and nonthermal X-rays \citep{Babazaki18, Yamane21}, suggesting the presence of a large quantity of accelerated multi-TeV electrons and a shock-amplified magnetic field. Here, we measure $\Gamma = 1.87\pm0.17$ and $\Sigma_{\Gamma} = (2.31\pm0.16)\times10^{-14}\,\si{erg.s^{-1}.cm^{-2}.arcmin^{-2}}$, yielding an integrated flux of $F_{\Gamma} = (2.29\pm0.15)\times10^{-12}\,\si{erg.s^{-1}.cm^{-2}}$. 
Our measured spectrum appears harder than that observed in the western shell with eROSITA \citep[$\Gamma = 2.4^{+0.2}_{-0.1}$;][]{Sasaki21}, and measurements based on other missions yielding similarly large photon indices \citep{Lopez20, Bamba04}. However, after conversion between energy bands, our estimate for the total nonthermal flux appears to fall into a reasonable range \citep{Lopez20}. 

\begin{figure}
\centering
\includegraphics[width=\linewidth]{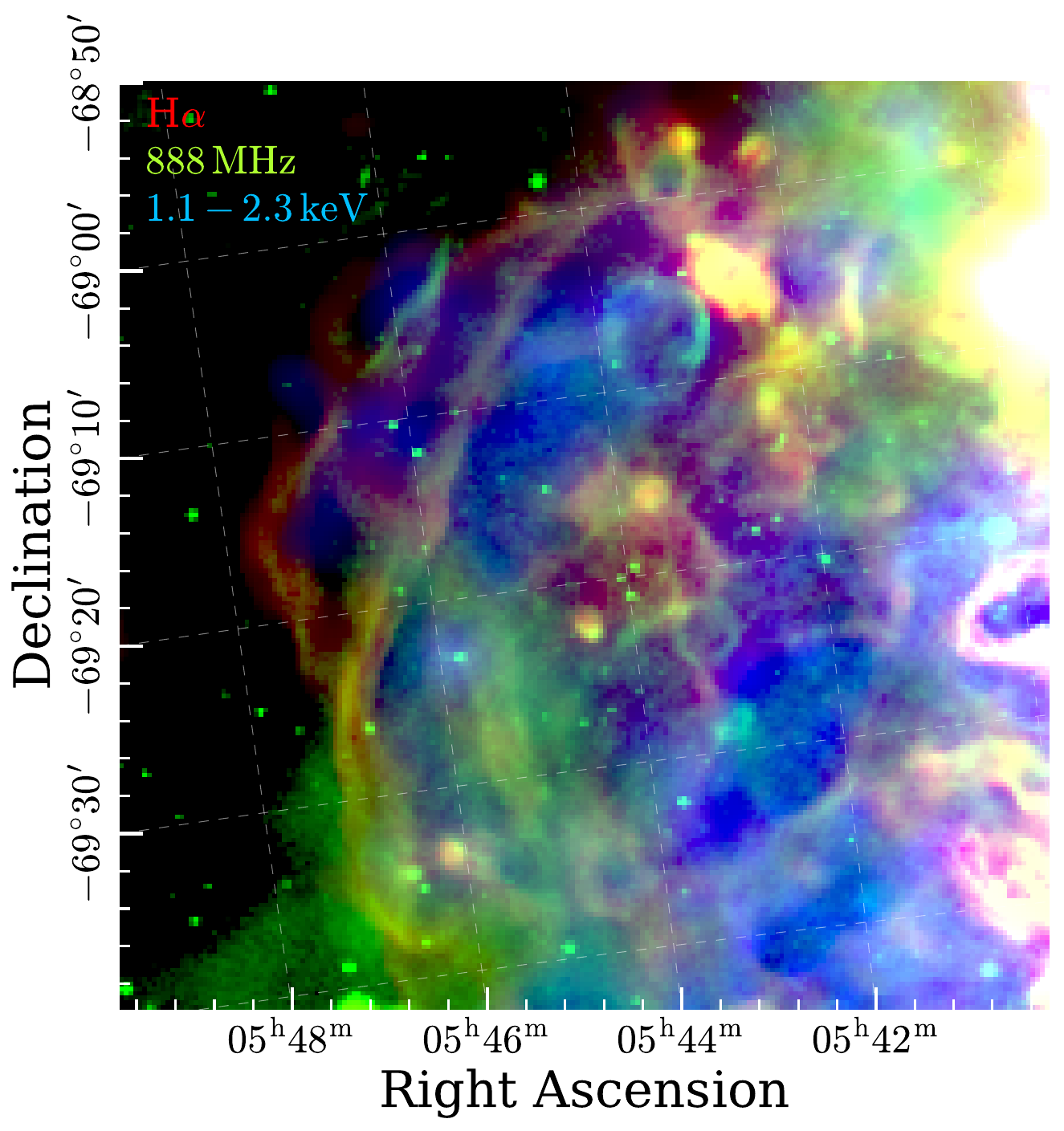} 
\caption{Composite false-color view of the eastern part of SGS LMC 2, showing H$\alpha$ line emission (red), ASKAP $888\,\si{MHz}$ radio continuum emission (green), and $1.1-2.3\,\si{keV}$ diffuse X-ray emission (blue). } 
\label{SGS2Image}
\end{figure}

Finally, we would like to discuss the possibility of nonthermal X-ray emission from the SGS LMC 2 \citep{Meaburn80}, located east of 30 Dor \citep[labelled as SGS 19 in ][]{Kim99}. Inspecting Fig.~\ref{SpecImage}, we note the presence of significant putative nonthermal emission fitted to the eastern part of LMC 2, which may be physical, but contamination by LMC X-1 appears likely. The X-ray emission of LMC 2 has been studied in the past \citep{Warth14a, Warth14b}, but the possibility of nonthermal emission was not considered.  
Figure \ref{SGS2Image} displays a composite view of LMC 2, incorporating MCELS H$\alpha$ emission \citep{Smith99},  ASKAP radio continuum emission at $888\,\si{MHz}$ \citep{Pennock21, Hotan21}, and our smoothed X-ray data in the ``hard'' $1.1-2.3\,\si{keV}$ band. Both radio and optical line emission trace the filamentary structure of the shell \citep{Meaburn80,Points99} in the north and east. In contrast, the X-ray emission 
appears to smoothly fill the inside of LMC 2, as reported already with {\it ROSAT} and \textit{XMM-Newton} \citep{Points00, Warth14b}. Given its large size, SGS LMC 2 is likely to be around $10^7\,\si{yr}$ old, but is continually energized by stellar winds and supernovae of young massive stars in the nearby star-forming regions \citep{Wang91b, Book08}, reaching a total thermal energy content around $10^{53}\,\si{erg}$ \citep{Warth14b}. 
In this context, the idea of the presence of diffuse TeV-energy electrons emitting X-ray synchrotron emission is interesting, given the results of our spectral fit, particularly since superbubbles, younger equivalents of SGSs, are considered efficient particle accelerators \citep{Parizot04}. The nondetection of the presumed $\gamma$-ray emission of these particles \citep{HESS15b} may be explained by their low surface brightness in comparison to other detected LMC sources.  

In order to investigate the origin of the putative nonthermal diffuse emission, it is necessary to quantify the contamination of the surroundings of LMC X-1 through straylight. Unfortunately, the average eRASS point spread function (PSF) is currently only calibrated out to $4\arcmin$ \citep{Merloni23}. In contrast, here, we are interested in distances up to $1^{\circ}$ from the source. \citet{Churazov23} attempted to construct a survey-average PSF model out to $3^{\circ}$. However, their work did not consider the effect of dust scattering halos around bright absorbed sources inside our Galaxy \citep[e.g.][]{Predehl95}, so that their model visibly overestimates the effective PSF for extragalactic sources, where dust scattering is not significant. 
Hence, we constructed our own unbiased PSF profile based on the nearby bright source LMC X-2 \citep[e.g.,][]{BonnetBidaud89}. To do this, we used the X-ray image and events in the $1.1-2.3\,\si{keV}$ band, integrated the counts in concentric annuli (starting at $30\arcsec$ to avoid the piled-up core), and fitted the PSF with an analytical model similar to that of \citet{Churazov21}. LMC X-2 is much better suited for the purpose of PSF calibration than LMC X-1 itself, due to its location far away from any contaminating diffuse emission. 
In order to apply the obtained model to our measurements, it is necessary to know the effective source flux in the relevant energy band, given the boundary conditions of the fit. In our case, this implies an incorrect physical model \citep[a power law rather than a disk black body;][]{Nowak01}. Furthermore, it implies an underestimated absorption, as the fitted $N_{\rm H}$ value in diffuse regions is driven by the soft thermal emission, not the contaminating hard straylight. 
In order to reproduce these conditions, we extracted a spectrum from an annulus (excising the central $30\arcsec$) centered on LMC X-1. We fitted this spectrum, fixing the column density to the representative value of $N_{\rm H} = 2.5\times10^{21}\,\si{cm^{-2}}$, and restricted the fitted spectral range to $1.0-2.3\,\si{keV}$. This latter choice was made to include only those energies where the straylight would be dominant over both thermal emission and particle background in our diffuse regions. 

\begin{figure}
\centering
\includegraphics[width=\linewidth]{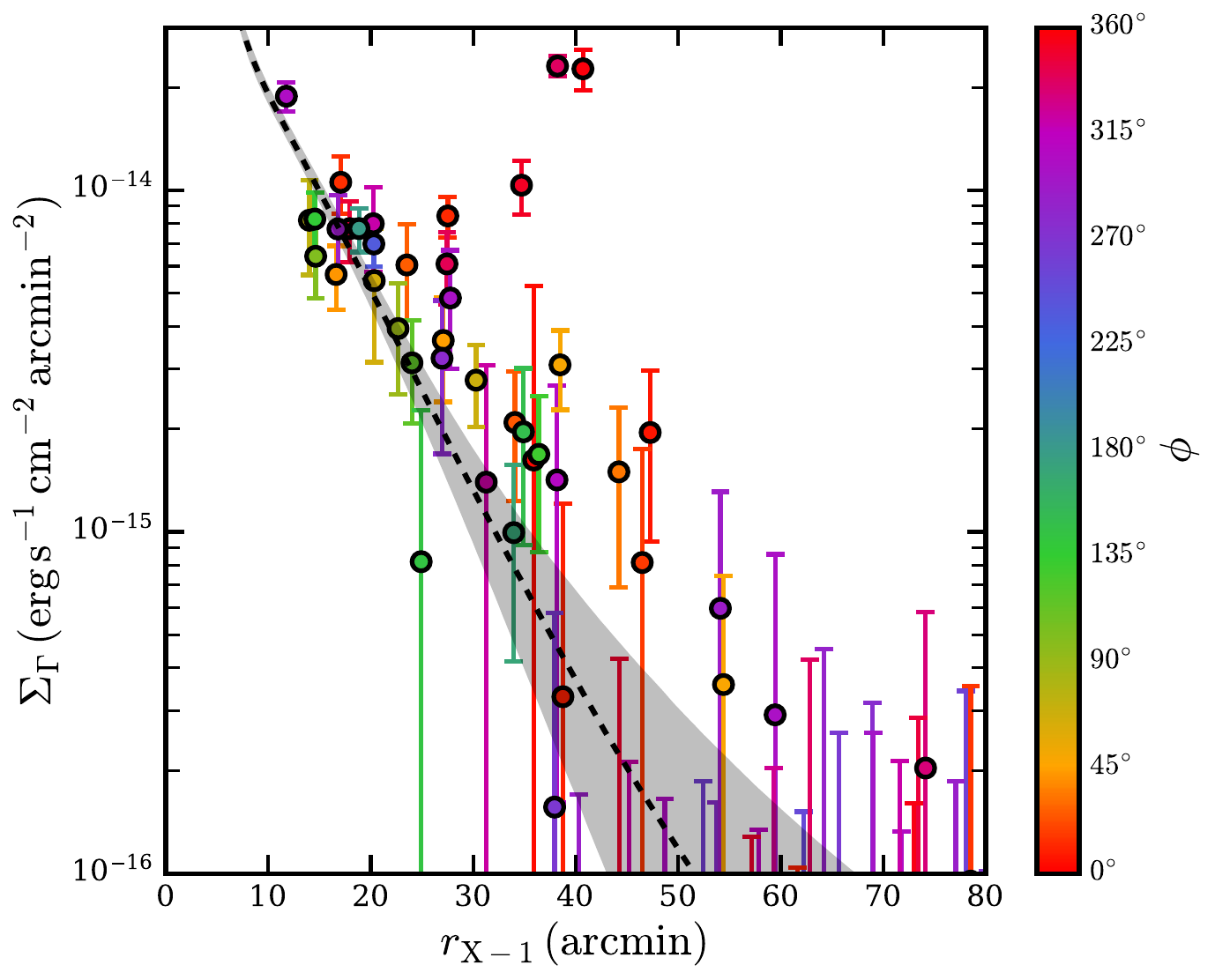} 
\caption{Comparison of measured nonthermal surface brightness in the $1.0-5.0\,\si{keV}$ band (as in Fig.~\ref{SpecImage}) against mean distance $r_{\rm X-1}$ of the region from LMC X-1. The marker color indicates the position angle $\phi$ (east of north) of each region with respect to LMC X-1. 
The dashed line indicates the expected flux level given the PSF profile, with the shaded region indicating its $1\sigma$ uncertainty.} 
\label{NonthermalPSF}
\end{figure}

By combining the resulting flux estimate of \mbox{$F_{X-1}=9.5\times10^{-10}\,\si{erg.s^{-1}.cm^{-2}}$} with our PSF model, we were able to estimate the expected radial profile of nonthermal emission from the contamination by LMC X-1 alone.       
This profile is displayed and compared to the measured surface brightness measurements in nearby regions in Fig.~\ref{NonthermalPSF}. Clearly, it captures the distance-dependent lower limit to the measured surface brightness well, considering the significant statistical uncertainties in both our fits and the PSF model. 
However, a few regions (at $r_{\rm X-1}\sim 40\arcmin$ and $\phi\sim0^{\circ}$) stand out significantly as regions of strong physical synchrotron emission, corresponding to R136 and 30 Dor C, as discussed above. 
The relevant regions of SGS LMC 2 are visible in our plot as orange-red markers at distances $r_{\rm X-1} \sim 25\arcmin - 50\arcmin$ from LMC X-1. A few regions appear to mildly exceed the background level expected from contamination alone (by up to $3\sigma$). However, it is unclear whether this can be considered a significant detection of physical synchrotron emission, given the systematics impacting both the determination of nonthermal flux in our fitted spectra, and our PSF profile. 
Unfortunately, even though shallow {\it XMM-Newton} observations of SGS LMC 2 exist \citep[see e.g.,][]{Knies21}, they are even more strongly affected by straylight arcs caused by LMC X-1, and hence do not provide more reliable constraints on the presence of diffuse nonthermal emission.
While uncertain, the detection of X-ray synchrotron emission inside LMC 2 would be an exciting finding. Even in the presence of an enhanced magnetic field, as seen in the nearby 30 Dor region \citep[e.g.,][]{Mao12, HESS15b, Kavanagh19}, this would require the presence of electrons at energies of $\gtrsim 10\,\si{TeV}$ to reproduce synchrotron photons with $\gtrsim1\,\si{keV}$. This, in turn, would imply ongoing or very recent particle acceleration in stellar winds or supernovae \citep{Parizot04}, given the limited lifetimes of leptonic cosmic rays under radiative losses.    

\section{Summary \label{Summary}}
In this work, we have analyzed the diffuse X-ray emission in the LMC based on the data set gathered during the full eROSITA all-sky survey (eRASS:5), using imaging and spectroscopy methods in a multiwavelength context. The imaging view of the hot ISM phase reveals a relatively smooth distribution of emission at $<0.7\,\si{keV}$ across the majority of the galaxy, consistent with a large filling factor of hot gas. In contrast, harder emission is strongly focussed to the southeast, surrounding the massive star-forming region 30 Dor.

By performing spatially resolved spectroscopy, and excluding point-like sources, we have decomposed the observed diffuse X-ray radiation into its different components in emission and absorption. While previous constraints on the energetics and mass of the hot gas in the LMC exist, our analysis for the first time allows for separating compact sources of thermal emission, such as SNRs, from diffuse emission from the ISM. For the latter, we measure an integrated luminosity of $L=1.9\times10^{38}\,\si{erg.s^{-1}}$, and a mean plasma temperature around $kT_{\rm mean} \approx 0.25\,\si{keV}$. We find typical electron densities for the hot ISM phase around $5\times10^{-3}\,\si{cm^{-3}}$, corresponding to a total hot-gas mass around $6\times10^6\,\si{M_{\odot}}$. While this mass is negligible compared to cooler phases, the pressure of the hot ISM is significant, being in the range $P/k \sim 10^4 - 10^5\,\si{K.cm^{-3}}$, as is the thermal energy content which we estimate to $9\times10^{54}\,\si{erg}$. 
Based on our energetic census of the hot ISM phase, we constructed characteristic cooling and heating timescales for the plasma, demonstrating that radiative cooling is at least a factor 20 slower than supernova heating of the plasma, implying other channels of energy loss in the hot phase.

While the typically used spectral model for thermal X-ray emission in the LMC consists of two CIE plasma components, we have also investigated alternative models: for instance, a continuous distribution of plasma temperatures typically leads to similarly good spectral fits as the discrete components, while implying a significantly larger amount of cool plasma and a significantly higher average density. 
In contrast, we find only weak evidence for NEI from recent shock heating, with most regions exhibiting shock ages $\gtrsim10^6\,\si{yr}$, apart from the core of the active \ion{H}{ii} region 30 Dor. 
Finally, we investigated the potential contribution of CX emission at hot-cold interfaces in the ISM. We found that, while a CX component cannot replace the CIE plasma component in our modelling without deteriorating the fit, a significant contribution to the X-ray emission on the order of $25\%$ cannot be excluded. 

Comparing the X-ray emission to that of cold gas and dust shows that, while the hot gas mostly remains inside the boundaries of the LMC disk, the hot and cold phases are anticorrelated on small scales. This, on one hand, manifests itself in a few ``holes'' in X-ray emission coincident with peaks in the distribution of cold ISM. 
On the other hand, the supergiant shells, large \ion{H}{i} cavities likely blown by massive stars, are smoothly filled with hot X-ray emitting gas. 
The distribution of absorbing foreground gas intrinsic to the LMC agrees very well between our spectral analysis and the emission of \ion{H}{i} gas, with a large absorbing column along the southeast rim, while the remainder of the galaxy is virtually unabsorbed in X-rays. 
The region which clearly stands out in this comparison is the ``X-ray spur'' \citep{Knies21}, a region of bright diffuse X-ray emission in the southeast, without an obvious origin from massive stellar energy input. For this region, while we do not find a significantly enhanced plasma temperature, we observe a pronounced peak in pressure of the hot ISM phase, as well as decreased $\alpha$-element abundances compared to the surroundings. Both these findings fit into the scenario in which the X-ray spur was caused by the tidally induced collision between different gas components, compressing and heating the diffuse gas.   

In contrast to the X-ray spur, regions further north, surrounding 30 Dor, were found to exhibit enhanced abundances of the light elements oxygen, neon, and magnesium relative to iron. Since iron is predominantly produced in type Ia supernovae, this $\alpha$-enhancement is likely a signature of the enrichment of the surrounding ISM by massive stars exploding in core-collapse supernovae.  
A possible signature of ISM enrichment through outflows from massive stars could be visible in SGS 17, north of 30 Dor, where a cavity in the cold ISM phase is filled with hot X-ray emitting plasma, and exhibits filamentary optical line emission.    
A particularly interesting feature visible in imaging is a collection of roughly coaligned short filaments west of the X-ray spur, emitting primarily in the $0.7-1.1\,\si{keV}$ band, and without multiwavelength counterpart. While their exact origin is unclear, we speculate that their common orientation may be related to the strong magnetic fields in this region of the LMC. 

Finally, our spectral analysis has also allowed us to test for the presence of diffuse nonthermal emission across the LMC. We find that the majority of the galaxy does not exhibit any significant X-ray synchrotron emission from energetic cosmic rays. The two secure exceptions are the core of 30 Dor, where the massive stellar cluster R136 is located, and the well-known superbubble 30 Dor C. 
Intriguingly, we also identify tentative signatures of X-ray synchrotron emission filling the supergiant shell LMC 2 east of 30 Dor. However, the low statistical significance, combined with the expected contamination from the X-ray binary LMC X-1, prevents a secure detection. 

Being the first part of a paper series, this work was primarily aimed at the spectral and morphological analysis of the X-ray emission of the LMC, with a qualitative interpretation in combination with multiwavelength data. 
In a subsequent work (Mayer et al., in prep.), we will perform a more thorough analysis of multiwavelength data to study the properties of the ISM. This will include the quantitative modelling of the energy input from underlying stellar populations, the metal enrichment of the hot ISM through supernovae, and the cross-correlation of the distribution of hot, warm, and cold gas in the ISM.

\begin{acknowledgements}
We are grateful to the anonymous referee for their constructive comments, which helped considerably improve this work. 
We would like to thank Yingjie Cheng and Shuinai Zhang for providing a spectral model for a plasma with continuous temperature distribution.
We would like to thank M.~G.~H.~Krause, K.~Nowak, and A. Zainab for fruitful discussions. 
M.G.F.M. acknowledges support from the Deutsche Forschungsgemeinschaft (DFG) through the grant MA 11073/1-1.  
M.S. acknowledges support from the DFG through the grants SA 2131/13-1, SA 2131/14-1, and SA 2131/15-1.
\\
This work is based on data from eROSITA, the soft X-ray instrument aboard SRG, a joint Russian-German science mission supported by the Russian Space Agency (Roskosmos), in the interests of the Russian Academy of Sciences represented by its Space Research Institute (IKI), and the Deutsches Zentrum f\"ur Luft- und Raumfahrt (DLR). The SRG spacecraft was built by Lavochkin Association (NPOL) and its subcontractors, and is operated by NPOL with support from the Max Planck Institute for Extraterrestrial Physics (MPE). The development and construction of the eROSITA X-ray instrument was led by MPE, with contributions from the Dr. Karl Remeis Observatory Bamberg \& ECAP (FAU Erlangen-Nuernberg), the University of Hamburg Observatory, the Leibniz Institute for Astrophysics Potsdam (AIP), and the Institute for Astronomy and Astrophysics of the University of T\"ubingen, with the support of DLR and the Max Planck Society. The Argelander Institute for Astronomy of the University of Bonn and the Ludwig Maximilians Universit\"at Munich also participated in the science preparation for eROSITA.
The eROSITA data shown here were processed using the eSASS software system developed by the German eROSITA consortium.
This scientific work uses data obtained from Inyarrimanha Ilgari Bundara, the CSIRO Murchison Radio-astronomy Observatory. We acknowledge the Wajarri Yamaji People as the Traditional Owners and native title holders of the Observatory site. CSIRO’s ASKAP radio telescope is part of the Australia Telescope National Facility (\url{https://ror.org/05qajvd42}). Operation of ASKAP and MWA is funded by the Australian Government with support from the National Collaborative Research Infrastructure Strategy. ASKAP and MWA use the resources of the Pawsey Supercomputing Research Centre. Establishment of ASKAP, Inyarrimanha Ilgari Bundara, the CSIRO Murchison Radio-astronomy Observatory and the Pawsey Supercomputing Research Centre are initiatives of the Australian Government, with support from the Government of Western Australia and the Science and Industry Endowment Fund.
\\
This research made use of Astropy,\footnote{\url{http://www.astropy.org}} a community-developed core Python package for Astronomy \citep{astropy:2013, astropy:2018}. Further, we acknowledge the use of the Python packages Matplotlib \citep{Hunter:2007}, SciPy \citep{SciPy}, and NumPy \citep{NumPy}. In particular, we acknowledge the use of the {\tt cubehelix} color map by \citet{Green11}.  
\\
\end{acknowledgements}
\vspace{-0.5cm}
\begingroup
\let\clearpage\relax

\bibliographystyle{aa} 
\bibliography{Citations} 

\endgroup

\end{document}